\documentclass{jpp}
\usepackage{graphicx}
\usepackage{epstopdf, epsfig}
\usepackage{hyperref}

\usepackage{amssymb}
\usepackage{natbib}
\usepackage{amsmath}
\usepackage{subfigure}
\usepackage{mathtools}

\shorttitle{Distribution functions for force-free current sheets}
\shortauthor{F. Wilson, T. Neukirch and O. Allanson}

\title{Collisionless distribution functions for force-free current sheets: using a pressure transformation to lower the plasma beta}

\author{F. Wilson\aff{1}
  \corresp{\email{fionaw237@gmail.com}},
  T. Neukirch\aff{1}
 \and O. Allanson\aff{1,2}}

\affiliation{\aff{1}School of Mathematics and Statistics, University of St Andrews, St Andrews, KY16 9SS, UK
\aff{2}Space and Atmospheric Electricity Group, Department of Meteorology, University of Reading, Reading, RG6 6BB, UK.}

\begin{document}

\maketitle

\begin{abstract}
So far, only one distribution function giving rise to a collisionless nonlinear force-free current sheet equilibrium allowing for a plasma beta less than one is known \citep{Allanson-2015, Allanson-2016}. This distribution function can only be expressed as an infinite series of Hermite functions with very slow convergence and this makes its practical use cumbersome. It is the purpose of this paper to present a general method that allows us to find distribution functions consisting of a finite number of terms (therefore easier to use in practice), but which still allow for current sheet equilibria that can, in principle, have an arbitrarily low plasma beta. The method involves using known solutions and transforming them into new solutions using transformations based on taking integer powers ($N$) of one component of the pressure tensor. The plasma beta of the current sheet corresponding to the transformed distribution functions can then, in principle, have values as low as $1/N$. We present the general form of the distribution functions for arbitrary $N$ and then, as a specific example, discuss the case for $N=2$ in detail.

%
\end{abstract}

\section{Introduction}
\label{sec:intro}
Force-free current sheets, for which
\begin{eqnarray}
\nabla\boldsymbol{\cdot}\textbf{B}&=&0\label{ff1},\\
\nabla\times\textbf{B}&=&\mu_0\textbf{j}\label{ff2},\\
\textbf{j}\times\textbf{B}&=&\textbf{0}\label{ff3},
\end{eqnarray}
are thought to form in the solar atmosphere and planetary magnetospheres and are, therefore, often used for modelling purposes (e.g. \cite{Bobrova-1979, Kivelson-1995, Marshbook,Tassi-2008, Wiegelmann-2012, priest_2014, Akcay-2016, Burgess-2016, Gingell-2017,Borissov-2017, Huang-2017, Lukin-2018}). Equations (\ref{ff1})-(\ref{ff3}) imply that the magnetic field, $\textbf{B}$, and the current density, $\textbf{j}$, are parallel to each other, i.e. that $\textbf{j}=\alpha(\textbf{r})\textbf{B}$. If $\alpha$ is constant, we have a linear force-free field, of which $\alpha=0$ (a potential field) is a special case. If $\alpha$ varies with position, we have a nonlinear force-free field.

In-situ spacecraft observations of approximately force-free current sheets in the magnetospheres of the Earth, Jupiter and Mars have been discussed by \cite{Panov-2011, Vasko-2014a, Artemyev-2014, Zelenyi-2016, Artemyev-2017a, Artemyev-2017b}, while observations of cylindrical force-free magnetic flux ropes in the Earth's magnetosphere have been discussed by \cite{Vinogradov-2016} (see \cite{Allanson-2016b} for a theoretical approach). The scales of these observed current sheets/flux ropes are kinetic and so, to initialise theoretical studies of the dynamics of such structures, we can use Vlasov-Maxwell (VM) equilibrium models. An important application of such models is to studies of collisionless magnetic reconnection. For examples of such studies using either exact or approximate VM equilibrium models for force-free current sheets, see \cite{Bobrova-2001,Nishimura-2003,Hesse-2005,Bowers-2007,Liu-2013, Guo-2014, Guo-2015, Zhou-2015, Wilson-2016, Guo-2016a, Guo-2016b, Fan-2016}. 

Since current sheets are strongly localised (spatially), they are often well described by one-dimensional (1D) VM equilibrium models.  Several authors have discussed 1D VM equilibrium models for linear force-free fields (e.g., \cite{Moratz-1966,Sestero-1967,Channell-1976,Correa-Restrepo-1993,Attico-1999,Bobrova-2001,Harrison-2009a}) and, in recent years, there has been considerable progress in finding exact analytical VM equilibrium distribution functions (DFs) for 1D nonlinear force-free fields. Examples include DFs found for the \textit{force-free Harris} current sheet,
\begin{equation}
\textbf{B}=B_0\left(\tanh(z/L), \mbox{sech}(z/L), 0\right),\label{Bffhs}
\end{equation}
(\cite{Harrison-2009b, Neukirch-2009, Wilson-2011, Stark-2012, Kolotkov-2015}) and for generalisations of this current sheet model (\cite{Abraham-Shrauner-2013, Wilson-2017}). For a discussion of ``semi-analytic'' DFs for a magnetic field that includes the force-free Harris sheet as a special case, we refer the reader to \cite{Dorville-2015}. We note that the analytical models referenced above all have vanishing $B_z$. For modelling environments such as the Earth's magnetosphere, however, it may be more appropriate to use models with a finite $B_z$, since observed current sheets usually have non-zero $B_z$ (see the observational studies referenced above). A finite $B_z$ has been taken into account in, e.g., the work by \cite{Artemyev-2011} and \cite{Vasko-2014a}. This can also change the stability properties of the system (see, e.g., \cite{schindlerbook} for a discussion).

In the exact equilibrium models discussed by, e.g., \cite{Channell-1976, Attico-1999, Harrison-2009b, Wilson-2011, Kolotkov-2015, Abraham-Shrauner-2013, Wilson-2017}, the plasma beta, $\beta_{pl}$, defined as the ratio of the plasma pressure and the magnetic pressure, is constrained to be at least one. This can be seen as a shortcoming of these models since, in the physical environments that might be modelled using force-free fields, the plasma pressure is often smaller than the magnetic pressure. 
Exceptions to this are the DFs for linear force-free collisionless current sheets presented by, e.g. \cite{Sestero-1967} and \cite{Bobrova-2001}, which in principle allow for $\beta_{pl} < 1$ \citep[see e.g. the discussion in][p.\ 763]{Bobrova-2001}.
We also note that there is always the possibility of adding a constant background magnetic field in order to lower the plasma beta, such as is often done for the Harris sheet. 
An example of particle-in-cell simulations of collisionless reconnection using a force-free field with an additional constant background field can be found in \cite{Hesse-2005}.
However, as pointed out by \cite{Harrison-2009a} for the Harris sheet, an additional constant guide field would not add any free energy to the system and a stronger constant guide field component might, in analogy to the Harris sheet with constant guide field case \citep[see e.g.][]{Pritchett-2004}, lead to a reduction of the reconnection rate.
Furthermore, the analytical methods used to derive the DFs above may not be suitable for the case of a force-free current sheet with an additional constant guide field. 

 \cite{Allanson-2015,Allanson-2016} managed to overcome the $\beta_{pl}\ge1$ problem by using a pressure transformation method (as suggested by \cite{Harrison-2009a}) to find DFs for the force-free Harris sheet, for which the plasma beta can be arbitrarily small. We note that, for the case of linear force-free collisionless current sheets, a similar pressure transformation 
links the DFs of \cite{Channell-1976} and \cite{Attico-1999} with those 
by, e.g. \cite{Sestero-1967} and \cite{Bobrova-2001} (see also \cite{Bobrova-2003}). As mentioned above, it is possible to have values of the plasma beta that are less than one in the linear force-free case with the latter DFs.

So far, the DF found by \cite{Allanson-2015, Allanson-2016} is the only case of an explicitly known analytical equilibrium DF allowing a nonlinear force-free collisionless current sheet with a plasma beta smaller than one. Furthermore, as discussed in some detail by \cite{Allanson-2016}, the practical use of this DF is problematic since it consists of infinite sums over Hermite functions, which, for example, give rise to issues with numerical convergence. This raises the question of whether other analytical DFs can be found that also allow for a plasma beta smaller than one, but which are easier to use in practice. It is the main purpose of this paper to present a method with which a class of DFs can be found that answers the above question positively. This method is again based on a pressure transformation that allows the plasma beta to become smaller than one. Instead of using an exponential pressure transformation as done by \cite{Allanson-2015, Allanson-2016}, we use a positive integer power $N$ of the pressure function as a transformation, which in principle allows the plasma beta to take values as low as $1/N$. 
We apply the method to the force-free Harris sheet and find that the DFs have a simpler form than those found by \cite{Allanson-2015, Allanson-2016}, since they only consist of a finite number of terms that are combinations of trigonometric and exponential functions. 
To illustrate the method, we give a comprehensive discussion of the case $N=2$, which has already been briefly mentioned by \cite{Neukirch-2017}.


The paper is laid out as follows; in Section \ref{sec:1dvm}, we discuss some background theory of 1D VM equilibria; in Section \ref{sec:pressure_transformation}, we discuss the pressure transformation we will use; in Section \ref{sec:quadratic}, we focus on the particular example of the transform that has already been briefly discussed by \cite{Neukirch-2017}, describing the methods for finding the DF in more detail, giving a lengthier discussion of some of the properties of the DF, and presenting some illustrative plots. We end with a summary and conclusions in Section \ref{sec:summary}.

\section{1D Vlasov-Maxwell equilibria}
\label{sec:1dvm}
In this section, we will briefly describe some of the relevant background theory of 1D VM equilibria. Following some previous work in this area (e.g. \cite{Harrison-2009a,Harrison-2009b,Neukirch-2009}), we assume that all quantities depend only on $z$, and that the magnetic field, $\textbf{B}=(B_x,B_y,0)$, can be written as $\textbf{B}=\nabla\times\textbf{A}$, where $\textbf{A}=(A_x, A_y, 0)$ is a vector potential. We also 
choose the parameters of our DFs in such a way
that the electric field vanishes, which is consistent with assuming that $\textbf{E}=-\nabla\phi$ for vanishing scalar potential, $\phi$ (i.e. strict neutrality). Under these assumptions, the VM equations reduce to solving Amp\`{e}re's law in the form
\begin{eqnarray}
\frac{\mathrm{d}^2A_x}{\mathrm{d}z^2}&=&-\mu_0\frac{\partial P_{zz}}{\partial A_x}\label{ampx},\\
\frac{\mathrm{d}^2A_y}{\mathrm{d}z^2}&=&-\mu_0\frac{\partial P_{zz}}{\partial A_y}\label{ampy},
\end{eqnarray}
for $P_{zz}$, which is the $zz$-component of the pressure tensor, defined by
\begin{equation}
P_{zz}(A_x, A_y)=\sum_sm_s\int v_z^2f_s(H_s, p_{xs}, p_{ys})\mathrm{d}^3v\label{pzz def}.
\end{equation} 
Note that we only consider the $zz$-component of the pressure tensor, since it is the component that is important for the force-balance of the 1D equilibrium (e.g. \cite{Harrison-2009a}). We assume that the DFs, $f_s$, are functions of the particle energy, $H_s=m_s(v_x^2+v_y^2+v_z^2)/2$, and the $x$- and $y$-components of the canonical momentum, $\textbf{p}_s=m_s\textbf{v}+q_s\textbf{A}$ (for $m_s$ the mass and $q_s$ the charge of species $s$, respectively), since these are known constants of motion for a time-independent system with spatial invariance in the $x$- and $y$-directions. Furthermore, following a method by \cite{Channell-1976} (see also \cite{Alpers-1969}), we assume that the DFs have an exponential dependence on $H_s$ and an arbitrary dependence on $p_{xs}$ and $p_{ys}$,
\begin{equation}
f_s=\frac{n_{0s}}{\left(\sqrt{2\pi}v_{th,s}\right)^3}e^{-\beta_sH_s}g_s\left(p_{xs}, p_{ys}\right),\label{fs_form}
\end{equation}
for which the pressure, $P_{zz}$, is given by 
\begin{eqnarray}
{P}_{zz}&=&\frac{\beta_e+\beta_i}{\beta_e\beta_i}\frac{n_{0s}}{2\pi m_s^2v_{th,s}^2}\nonumber\\
&{}&\times\int_{-\infty}^{\infty}\int_{-\infty}^{\infty}\exp\left[-\frac{\beta_s}{2m_s}\left((p_{xs}-q_sA_x)^2+(p_{ys}-q_sA_y)^2\right)\right]g_s(p_{xs}, p_{ys})\mathrm{d}p_{xs}\mathrm{d}p_{ys},\label{inverse_prob}\nonumber\\
\end{eqnarray}
where $n_{0s}$ is a constant with the dimension of number density, $\beta_s=\left(k_BT_s\right)^{-1}$ and $v_{th,s}=\left(\beta_sm_s\right)^{-1/2}$, for constant temperature $T_s$. The problem now consists of finding the unknown function(s) $g_s$ that are consistent with the specified magnetic field profile through Equations (\ref{ampx}), (\ref{ampy}) and (\ref{inverse_prob}). This is an example of an ``inverse problem'', and can be solved by, e.g., using Weierstrass transforms (e.g. \cite{Allanson-2018}).

As discussed by, e.g., \cite{Harrison-2009a}, the 1D VM equilibrium problem as described above is equivalent to the problem of determining the motion of a particle in a conservative potential, with $\mu_0P_{zz}(A_x,A_y)$ playing the role of the potential, ($A_x,A_y$) representing the position of the particle, and $z$ playing the role of time. For a 1D force-free magnetic field satisfying Equations (\ref{ff1})-(\ref{ff3}), a necessary condition for a VM equilibrium is that the trajectory $(A_x(z), A_y(z))$ is a contour of the pressure (i.e. the potential) $P_{zz}(A_x,A_y)$. In Section \ref{sec:pressure_transformation}, we will make use of this condition when discussing the pressure transformation method.

\section{Pressure transformation}
\label{sec:pressure_transformation}

For a given force-free magnetic field profile, the pressure function $P_{zz}(A_x,A_y)$ consistent with this magnetic field profile is not unique. As discussed by  \citet{Harrison-2009a}, given a pressure $P_{zz}$ that allows a force-free solution ($A_{x, ff}(z), A_{y, ff}(z)$), we can construct a new pressure function, $\bar{P}_{zz}$, for the same magnetic field by using the transformation
\begin{equation}
\bar{P}_{zz}=\frac{F(P_{zz})}{F'\left(P_{ff}\right)},\label{transformation}
\end{equation}
where $P_{ff}$ is the constant value of $P_{zz}$ evaluated on the force-free contour ($A_{x, ff}(z), A_{y, ff}(z)$), and $F$ is an arbitrary differentiable function, chosen so that the right-hand side of Equation (\ref{transformation}) is positive. The transformation works by deforming the pressure surface $P_{zz}(A_x, A_y)$ such that (a) the trajectory ($A_{x, ff}(z),A_{y, ff}(z)$) still corresponds to a contour of the resulting pressure surface $\bar{P}_{zz}$ and (b) the following relations hold on the force-free contour;
\begin{equation}
\frac{\partial \bar{P}_{zz}}{\partial A_x} = \frac{\partial {P}_{zz}}{\partial A_x}, \quad\frac{\partial \bar{P}_{zz}}{\partial A_y}=\frac{\partial {P}_{zz}}{\partial A_y}.
\end{equation}
These conditions ensure that the trajectory ($A_{x, ff}(z), A_{y, ff}(z)$) still corresponds to a force-free solution.

The transformation (\ref{transformation}) has previously been used by \cite{Allanson-2015, Allanson-2016}, by taking $F(P_{zz})=\exp\left[(P_{zz}-P_{ff})/P_{0}\right]$ (with $P_{0}$ a positive constant), to find DFs for the force-free Harris sheet (Equation (\ref{Bffhs})) in terms of infinite sums of Hermite polynomials. In this case, it is possible to choose an arbitrarily low plasma beta, compared with other work on DFs for the force-free Harris sheet, where the plasma beta is constrained to be greater than one \citep{Harrison-2009b, Neukirch-2009, Wilson-2011, Abraham-Shrauner-2013, Kolotkov-2015, Wilson-2017}. We note that, throughout this work, we define the plasma beta as $\beta_{pl}=P_{zz}/(B^2/(2\mu_0))$, instead of the more conventional definition given by $\bar\beta_{pl}=p/(B^2/(2\mu_0))$, where $p=(P_{xx}+P_{yy}+P_{zz})/3$. For a discussion of how the pressure transformation (\ref{transformation}) affects $\bar\beta_{pl}$ in the $N=2$ case, see Appendix \ref{sec:full_beta}.

As discussed by \cite{Allanson-2015}, the pressure function resulting from the DF for a linear force-free current sheet given in the work 
by \cite{Sestero-1967} (see also \cite{Bobrova-2001} and \cite{Bobrova-2003}) could be regarded as resulting from an exponential pressure transformation
(although, obviously, the transformation is not used in the original papers), because the pressure function used in these references is the function that results from taking the exponential of the pressure function discussed by \cite{Channell-1976} and \cite{Attico-1999}. In the cases considered by \cite{Channell-1976} and \cite{Attico-1999}, the plasma beta is constrained to be at least one, due to the summative nature of the pressure function (of the form $P_{zz}=P_{1}(A_x)+P_2(A_y)$). This is straightforward to show and, as discussed by \cite{Allanson-thesis}, it can be shown that, for certain types of force-free magnetic fields, the plasma beta is constrained to be at least one whenever a summative pressure function is assumed (see Appendix \ref{sec:summative} for further details). In the cases considered by \cite{Sestero-1967, Bobrova-2001, Bobrova-2003}, the plasma beta can be smaller than one, and so the exponential transformation has allowed for a reduction in the plasma beta, in a similar way to the nonlinear force-free Harris sheet case considered by \cite{Allanson-2015, Allanson-2016}. Essentially, it appears that if the transformation is such that the original summative pressure function is transformed to a pressure function that has a multiplicative part, i.e. a function of the form $\bar{P}_1(A_x)\bar{P}_2(A_y)$, then it is possible to achieve a plasma beta of lower than one. This might not always be true, however, depending on the particular transformation used, but it works for the exponential transformation (for the linear force-free case and the force-free Harris sheet), and for the transformation that we consider later in this paper.

One problem of the low-beta DFs found by \cite{Allanson-2015, Allanson-2016} is that they can be difficult to use practically, due to slow convergence of the infinite sums over Hermite functions that they involve. 
It is the main aim of this paper to use the general transformation property (\ref{transformation}) to find low-beta DFs with a simpler form. To achieve this aim, we will use the function
\begin{equation}
F(P_{zz}) = P_{zz}^N,\label{new F}
\end{equation}
where $N>1$, for the transformation. 
%

We start by illustrating why this transformation can give rise to $\beta_{pl}<1$, for a general $N$, on the macroscopic level. 
Using the function (\ref{new F}) in the transformation (\ref{transformation}) gives the transformed pressure as
\begin{equation}
\bar{P}_{zz}=\frac{1}{N}P_{ff}^{1-N}P_{zz}^N\label{pbar},
\end{equation}
and so $\bar{P}_{zz}$ evaluated on the force-free contour is given by
\begin{equation}
\bar{P}_{ff}=\frac{P_{ff}}{N}.
\end{equation}
This choice of $F$, therefore, allows us to find a pressure function that will give rise to a lower plasma beta than for the original $P_{zz}$. 
Increasing $N$ will make $\beta_{pl}$ as low as required.

We will now apply this transformation to the pressure function for the force-free Harris sheet found by \cite{Harrison-2009b, Neukirch-2009}, which is given by
\begin{equation}
P_{zz, hn} = \frac{B_0^2}{2\mu_0}\left[\frac{1}{2}\cos\left(\frac{2A_x}{B_0L}\right)+\exp\left(\frac{2A_y}{B_0L}\right)+b\right],\label{h&n p}
\end{equation}
as a function of $A_x$ and $A_y$.
On the force-free contour, we have 
\begin{equation}
P_{ff, hn} = \frac{B_0^2}{2\mu_0}\left(1/2+b\right)\label{hn_ff},
\end{equation}
which can be seen by substituting in the vector potential components
\begin{eqnarray}
A_x&=&2B_0L\tan^{-1}\left(e^{z/L}\right),\\
A_y&=&-B_0L\ln\left(\cosh(z/L)\right),
\end{eqnarray}
and by using the fact that
\begin{equation}
\cos\left(4\tan^{-1}\left(e^{z/L}\right)\right)=1-2\mbox{sech}^2(z/L),\label{cos4}
\end{equation}
(e.g. \citet{fionawilson-thesis}). 
The plasma beta for this case is then given by
\begin{equation}
\beta_{pl}=\frac{{P}_{ff, hn}}{(B_0^2/(2\mu_0))}=1/2+b,
\end{equation}
where we note that the magnetic pressure for the force-free Harris sheet is equal to $(B_x^2+B_y^2)/(2\mu_0)=B_0^2/(2\mu_0)$. The parameter $b$ is constrained to be at least $1/2$, so that any DFs calculated from the pressure function (\ref{h&n p}) are positive over the whole phase space (\cite{Harrison-2009b, Neukirch-2009, Wilson-2011, Stark-2012, Kolotkov-2015, Abraham-Shrauner-2013, Wilson-2017}). This means that $\beta_{pl}\ge1$ for these models, as mentioned above. 

Equations (\ref{h&n p}) and (\ref{hn_ff}) can now be substituted into Equation (\ref{pbar}), to construct a new pressure function for the force-free Harris sheet, given by 
\begin{equation}
\bar{P}_{zz}=\frac{B_0^2}{2\mu_0}\frac{\left(1/2+{b}\right)^{1-N}}{N}\left[\frac{1}{2}\cos\left(\frac{2A_x}{B_0L}\right)+\exp\left(\frac{2A_y}{B_0L}\right)+b\right]^N.\label{pzz_macro}
\end{equation}
On the force-free contour, this transformed pressure function has the value
\begin{equation}
\bar{P}_{ff}=\frac{B_0^2}{2\mu_0}\frac{1/2+{b}}{N},
\end{equation}
and so the plasma beta for the force-free solution is given by
\begin{equation}
\beta_{pl}=\frac{1}{N}(1/2+{b})\label{transformed beta},
\end{equation}
which is clearly less than that found from the pressure function (\ref{h&n p}) when $N>1$. 
The next step is to find the DFs for the transformed pressure, which is done by substituting the transformed pressure (\ref{pzz_macro}) into Equation (\ref{inverse_prob}), and solving for the unknown function(s), $g_s$. We state the general expression for the DFs in Appendix \ref{sec:general dfs}. The conditions on the parameters of the new DFs must be such that the new DFs are positive over the whole phase space. This will give a lower bound for the plasma beta, which could in principle be higher than $1/N$ depending on other parameter values. 
As an illustration, we present the detailed calculation for the special case of the quadratic  transformation ($N=2$) in Section \ref{sec:quadratic}, 
which has already been briefly discussed by \cite{Neukirch-2017}. 
In this case, we find that, for certain parameter values, 
the lower bound on the plasma beta can indeed be reduced to $1/2$, i.e. 
by a factor of two compared with the solution found by \cite{Harrison-2009b}.

\section{The quadratic case (\texorpdfstring{$N=2$}{lg})}

\label{sec:quadratic}
In this section, we will present (as an illustration of the general method) the quadratic transformation of the force-free Harris sheet case (i.e. we will set $N=2$ in Equation (\ref{pbar})). 
We are considering this case in detail as a compromise between presenting an explicit example for the workings of the transformation, and keeping the calculations reasonably short (for larger values of $N$ the calculations become lengthier and more involved). The main difficulty comes from finding conditions under which the DF is always positive over the whole phase space.  
A very brief discussion of the quadratic case has already been given in the review paper by \cite{Neukirch-2017}, but we will give a much more detailed and systematic account of the calculation and the properties of the DFs.

\subsection{Pressure transformation and deriving the DF}
\label{sec:quad_p}
For the quadratic case, the transformed pressure is given by
\begin{equation}
\begin{split}
\bar{P}_{zz}= \frac{B_0^2}{2\mu_0}\left(1+2{b}\right)^{-1}\Bigg[\frac{1}{8}\cos\left(\frac{4A_x}{B_0L}\right)&+\exp\left(\frac{4A_y}{B_0L}\right)+\cos\left(\frac{2A_x}{B_0L}\right)\exp\left(\frac{2A_y}{B_0L}\right) \\
 &+{b}\cos\left(\frac{2A_x}{B_0L}\right)+2{b}\exp\left(\frac{2A_y}{B_0L}\right)+{b}^2+\frac{1}{8}\Bigg]\label{quad p},
\end{split}
\end{equation}
and on the force-free contour this function has the value
\begin{equation}
\bar{P}_{ff}=\frac{B_0^2}{2\mu_0}\frac{(1/2+{b})}{2}\label{pff_quad},
\end{equation} 
giving a plasma beta equal to
\begin{equation}
\beta_{pl}=\frac{\bar{P}_{ff}}{(B_0^2/(2\mu_0))}=\frac{1}{2}\left(\frac{1}{2}+{b}\right).\label{quad_beta}
\end{equation}
The lower limit of $\beta_{pl}$ depends on the lower limit of ${b}$. As discussed in Section \ref{sec:pressure_transformation}, ${b}$ has a lower bound of $1/2$ (required for positivity of the DFs discussed by \cite{Harrison-2009b}); based on this fact alone, the smallest obtainable value of the plasma beta is $1/2$, meaning that the quadratic transformation allows us to reduce the lower limit on $\beta_{pl}$ by a factor of two (compared with the work by \cite{Harrison-2009b, Neukirch-2009, Wilson-2011, Stark-2012, Kolotkov-2015, Abraham-Shrauner-2013, Wilson-2017}). We must emphasise, however, that it cannot be assumed \textit{a priori} that the new DFs resulting from the transformed pressure will be positive over the whole phase space, and so we must find conditions on the parameters such that this is the case (this could, in principle, change the lower bound on $b$). In the present case ($N=2$), the lower bound on $b$ is still $1/2$, depending on other parameter values, which will be discussed further in Section \ref{subsec:positivity}. The lower bound on the plasma beta for the $N=2$ case is, therefore, $1/2$. 

The transformed pressure function (\ref{quad p}) can be used in Equation (\ref{inverse_prob}), the solution of which will give new DFs for the force-free Harris sheet. Using the fact that cosine and exponential functions are eigenfunctions of the Weierstrass transform (e.g. \cite{Wolf-1977}), we can immediately write the DFs in the form
\begin{eqnarray}
f_s(H_s, p_{xs}, p_{ys})&=&\frac{n_{0s}}{\left(\sqrt{2\pi}v_{th,s}\right)^3}e^{-\beta_sH_s}\Big[a_{1s}\cos(2\beta_su_{xs}p_{xs})+a_{2s}\exp(2\beta_su_{ys}p_{ys})\nonumber\\
&{}&+a_{3s}\cos(\beta_su_{xs}p_{xs})\exp(\beta_su_{ys}p_{ys})+a_{4s}\cos(\beta_su_{xs}p_{xs})\nonumber\\
&{}&+a_{5s}\exp(\beta_su_{ys}p_{ys})+a_{6s}\Big],\nonumber\\
\label{quad_df}
\end{eqnarray}
where $u_{xs}$ and $u_{ys}$ are constants with the dimension of velocity, and the terms $a_{1s}$ to $a_{6s}$ are dimensionless constants. As mentioned in Section \ref{sec:1dvm}, we assume strict neutrality, i.e. that $n_i(A_x,A_y)=n_e(A_x,A_y)$. We must also ensure consistency between the macroscopic and microscopic parameters of the equilibrium, which involves ensuring that the macroscopic expression  (\ref{quad p}) for $\bar{P}_{zz}$ matches with the microscopic expression calculated from the $v_z^2$ moment of the DF (\ref{quad_df}). These requirements give rise to the following conditions,
\begin{eqnarray}
-e\beta_e\vert u_{xe}\vert=e\beta_i\vert u_{xi}\vert&=&\frac{2}{B_0L}=-e\beta_eu_{ye}=e\beta_iu_{yi}\Rightarrow \vert u_{xs}\vert=u_{ys}\label{uxs_uys},\\
n_{0e}&=&n_{0i}=n_0,\\
n_0\frac{\beta_e+\beta_i}{\beta_e\beta_i}&=&\frac{B_0^2}{2\mu_0}({1+2{b}})^{-1}\label{B0},\\
a_{1e}\exp\left(-\frac{2u_{xe}^2}{v_{th,e}^2}\right)&=&a_{1i}\exp\left(-\frac{2u_{xi}^2}{v_{th,i}^2}\right)=\frac{1}{8},\\
a_{2e}\exp\left(\frac{2u_{ye}^2}{v_{th,e}^2}\right)&=&a_{2i}\exp\left(\frac{2u_{yi}^2}{v_{th,i}^2}\right)=1,\\
a_{3e}&=&a_{3i}=1,\\
a_{4e}\exp\left(-\frac{u_{xe}^2}{2v_{th,e}^2}\right)&=&a_{4i}\exp\left(-\frac{u_{xi}^2}{2v_{th,i}^2}\right)={b},\\
a_{5e}\exp\left(\frac{u_{ye}^2}{2v_{th,e}^2}\right)&=&a_{5i}\exp\left(\frac{u_{yi}^2}{2v_{th,i}^2}\right)=2{b},\\
a_{6e}&=&a_{6i}={b}^2+\frac{1}{8}\label{a6s}.
\end{eqnarray}
These relations between the electron and ion parameters result from our use of Channell's method \citep{Channell-1976}, where strict neutrality ($n_i(A_x,A_y)=n_e(A_x,A_y)$) is imposed to make analytical progress, leading to $\phi=0$  in the quasineutrality condition. However, if we assume that $\phi=0$ from the start and allow only one of the species to have a DF of the form (\ref{quad_df}), with the other species having a simple Maxwellian DF providing a constant neutralising density, some of the conditions linking the electron and ion DF parameters could be relaxed.

Using the relations above, the DF can be expressed as
\begin{eqnarray}
f_s&=&\frac{n_{0}}{\left(\sqrt{2\pi}v_{th,s}\right)^3}e^{-\beta_sH_s}\Bigg[\frac{1}{8}e^{2\bar{u}_{xs}^2}\cos(2\beta_su_{xs}p_{xs})+e^{-2\bar{u}_{ys}^2}\exp(2\beta_su_{ys}p_{ys})\nonumber\\
&{}&+\cos(\beta_su_{xs}p_{xs})\exp(\beta_su_{ys}p_{ys})+{b}e^{\bar{u}_{xs}^2/2}\cos(\beta_su_{xs}p_{xs})\nonumber\\
&{}&+2{b}e^{-\bar{u}_{ys}^2/2}\exp(\beta_su_{ys}p_{ys})+{b}^2+\frac{1}{8}\Bigg]\label{quad_df2},\nonumber\\
\end{eqnarray}
where $\bar{u}_{xs}=u_{xs}/v_{th,s}$ and $\bar{u}_{ys}=u_{ys}/v_{th,s}$ (note that $u_{xs}^2=u_{ys}^2$ through Equation (\ref{uxs_uys})).

There are five free parameters that we need to specify to fully describe the equilibrium. For example, from Equations (\ref{uxs_uys})-(\ref{a6s}), we see that all of the parameters can be calculated if we specify $n_0$, $\beta_e$, $\beta_i$, $u_{xs}$ (for either ions or electrons) and ${b}$ (i.e. $\beta_{pl}$). By using Equations (\ref{quad_beta}), (\ref{uxs_uys}) and (\ref{B0}), we can express the current sheet half-width $L$ as
\begin{equation}
L=\left(\frac{\beta_e+\beta_i}{2\mu_0e^2\beta_e\beta_in_0\beta_{pl}(u_{yi}-u_{ye})^2}\right)^{1/2}\label{halfwidth}.
\end{equation}

For fixed values of $\beta_e$, $\beta_i$, $u_{xs}$ (for either ions or electrons) and $n_0$, therefore, increasing $\beta_{pl}$ results in a thinning of the current sheet. This is due to the fact that raising the number density, $n$ (Equation (\ref{quad_ns})), results in a higher plasma beta (when $n_0$ is fixed) and, for a larger number density, there are more current carrying particles available to produce the current density, $\textbf{j}$, meaning that the thickness of the current sheet can reduce. A similar conclusion was reached by \cite{Allanson-2015}. Note, however, that our parameter $u_{xs}$ is different from the parameter $u_s$ in the work by  \cite{Allanson-2015} - in that case $u_s$ is defined as the amplitude of the $x$- and $y$-components of the bulk-flow velocity, which in our case is equal to $u_{xs}/\beta_{pl}$. 

\subsection{Moments of the DF}
\label{subsec:moments}
The density, $n_s$, and components of the bulk-flow velocity, $\textbf{V}_s$, can be calculated from the DF (\ref{quad_df2}), 
and are given by
\begin{eqnarray}
n&=&n_0\left(1/2+{b}\right)^2\label{quad_ns},\\
{V_{xs}}&=&\frac{2u_{xs}}{(1/2+{b})}\sinh(z/L)\mbox{sech}^2(z/L)=\frac{u_{xs}}{\beta_{pl}}\sinh(z/L)\mbox{sech}^2(z/L)\label{vxi},\\
{V_{ys}}&=&\frac{2u_{ys}}{(1/2+{b})}\mbox{sech}^2(z/L)=\frac{u_{ys}}{\beta_{pl}}\mbox{sech}^2(z/L)\label{vyi}.
\end{eqnarray}
Note that the $z$-component of $\textbf{V}_s$ vanishes, since the DF (\ref{quad_df2}) is a stationary Maxwellian in this direction. 
The components of the current density are then given by
\begin{eqnarray}
j_x&=&en_0(1+2{b})(u_{xi}-u_{xe})\sinh(z/L)\mbox{sech}^2(z/L),\\
j_y&=&en_0(1+2{b})(u_{yi}-u_{ye})\mbox{sech}^2(z/L).
\end{eqnarray}
Using Equations (\ref{uxs_uys}) and (\ref{B0}), we can show that these expressions are equivalent to the expressions for the current density components in terms of the macroscopic parameters.

\subsection{Positivity of the DF and implications for the plasma beta and other parameters}
\label{subsec:positivity}

For a physically reasonable solution, we must ensure that the DF (\ref{quad_df2}) is positive over the whole phase space. In Appendix \ref{app:gs+}, we show that this requires us to choose the parameter ${b}$ such that
\begin{equation}
{b}\ge\frac{e^{\bar{u}_{xs}^2/2}}{2\sqrt{2}}\left(e^{\bar{u}_{xs}^2}+1\right)^{1/2},\label{g+cond}
\end{equation}
which must be true for both ions and electrons. The parameters $\bar{u}_{xi}$ and $\bar{u}_{xe}$ are related (through Equation (\ref{uxs_uys})) by
\begin{equation}
\bar{u}_{xi}^2=\frac{T_i}{T_e}\frac{m_i}{m_e}\bar{u}_{xe}^2\label{uxe_uxi},
\end{equation} 
and so, generally, $\bar{u}_{xi}^2$ will be much larger than $\bar{u}_{xe}^2$. This means that, if we choose ${b}$  such that the ion DF is positive, then the electron DF will also be positive, since the condition on ${b}$ is more strict for the ions than for the electrons.
From Equation (\ref{quad_beta}), we see that the value of the plasma beta also depends on the choice of $b$. For a ``low-beta'' case, which we take here to mean $\beta_{pl}<1$, we need to choose ${b}<3/2$, noting that the lower bound on ${b}$ for a positive DF depends on the value of $\bar{u}_{xs}^2$ through Equation (\ref{g+cond}). By plotting the quadratic function ${b}^2=e^{\bar{u}_{xs}^2}(e^{\bar{u}_{xs}^2}+1)/8$, we see that if we choose $\bar{u}_{xs}^2\le 1.3$ (approximately) then ${b}<3/2$ is permitted, and so we can obtain $\beta_{pl}<1$. This means that, if we require both $\bar{u}_{xi}^2$ and $\bar{u}_{xe}^2$ less than about $1.3$ for $\beta_{pl}<1$, we must take 
\begin{equation}
\bar{u}_{xe}^2\le1.3\times\left(\frac{T_i}{T_e}\frac{m_i}{m_e}\right)^{-1},\label{uxe_val}
\end{equation}
i.e. $\bar{u}_{xe}$ must be very small unless $T_e>>T_i$. 

As discussed in Section \ref{sec:quad_p}, the lowest attainable value of $\beta_{pl}$ in this case is $1/2$, obtained when ${b}=1/2$. From Equation (\ref{g+cond}), we see that this condition is consistent with the positivity condition when $u_{xs}=0$. For non-zero values of $u_{xs}$, however, the lower bound on ${b}$ will be higher, resulting in $\beta_{pl}>1/2$.

To illustrate allowable parameter sets, e.g., for use in simulations, we can write the current sheet half-width normalised to the ion inertial length, $d_i=c\left(ne^2/(\epsilon_0m_i)\right)^{-1/2}$, as
\begin{equation}
\frac{L}{d_i}=\left(\frac{1/2+{b}}{\left(1+T_e/T_i\right)\bar{u}_{xi}^2}\right)^{1/2},\label{L/di}
\end{equation}
where we have used the full density, $n$, given by Equation (\ref{quad_ns}), to define the ion inertial length.
Combining Equation (\ref{L/di}) with the positivity condition (\ref{g+cond}) then gives
\begin{equation}
\frac{L}{d_i}\ge\left(\frac{\sqrt{2}+e^{\bar{u}_{xi}^2/2}(e^{\bar{u}_{xi}^2}+1)^{1/2}}{2\sqrt{2}\left(1+T_e/T_i\right)\bar{u}_{xi}^2}\right)^{1/2},
\end{equation}
where we have used the positivity condition for the ion DF, since this gives a stricter condition on ${b}$ than that for the electron DF, as discussed above. 

As an example, if we take $\bar{u}_{xi}^2=1.2$ and $T_e/T_i=1.0$, we get a minimum value of $L/d_i=0.87$, which illustrates that, despite the restrictions of the positivity condition (\ref{g+cond}), we can choose reasonable parameter sets for which $\beta_{pl}<1$.

\subsection{Example plots of the DF}

\label{sec:df_plots}

In this section, we will show some illustrative plots of the quadratic DF (\ref{quad_df2}) for different parameter values. Figures \ref{fig:quad_df} and \ref{fig:quad_df2} show plots of the ion DF in the $v_x$-direction (with $v_y=0$) for various values of $u_{xi}/v_{th,i}$, for $z=0$ (Figure \ref{fig:quad_df}) and $z=0.5$ (Figure \ref{fig:quad_df2}). In each case, we have adjusted ${b}$ according to the condition (\ref{g+cond}) for positivity of the DF, and hence the plasma beta is different in each plot. The DFs are normalised to have a maximum value of one in each case.  Note that, for negative values of $z$, the DF has the opposite symmetry from that for positive $z$ values with respect to $v_x=0$ (i.e. the DF is symmetric under the transformation $v_x\to-v_x$, $z\to-z$) and so we only show plots for a positive value of $z$ ($z=0.5$) to illustrate the behaviour of the DF away from $z=0$.   

We see that, for small values of $u_{xi}/v_{th,i}$ (and hence $\beta_{pl}$), the DF is single-peaked but, as $u_{xi}/v_{th,i}$ is increased, multiple maxima eventually appear in the DF, and these are more pronounced at $z=0$ than at $z=0.5$. Due to the condition (\ref{g+cond}) on $b$, however, these multiple maxima appear for $\beta_{pl}$ values only modestly below one. For the parameter values considered by \cite{Neukirch-2017}, only single maxima were found. Since this work was carried out, however, we have found that we can take less restrictive values of $u_{xi}^2/v_{th,i}^2$ and still have positive DFs over the whole phase space. We also note that the quadratic DF can have more pronounced multiple maxima for higher values of $u_{xi}/v_{th,i}$. We do not focus on such cases, however, since they occur at increasingly high values of the plasma beta due to the condition (\ref{g+cond}). We have only plotted ion DFs here, since the electron DFs are single-peaked for the given parameter values (due to Equation (\ref{uxe_uxi}), the values of $u_{xe}^2/v_{th,e}^2$ will be very small unless we take a very large value of $T_e/T_i$).

Figures \ref{fig:quad_df_diff} ($z=0$) and \ref{fig:quad_df_diff2} ($z=0.5$) show, for the same values of $u_{xi}/v_{th,i}$ and $\beta_{pl}$ as in Figures \ref{fig:quad_df} and \ref{fig:quad_df2}, plots of the difference between the quadratic ion DF and the Maxwellian DF given by
\begin{equation}
f_{M,i}=\frac{n_{0,M}}{\left(\sqrt{2\pi}v_{th,s}\right)^3}\exp\left[-\frac{(\textbf{v}-\textbf{V}_s)}{2v_{th,s}^2}^2\right],\label{max_df}
\end{equation}
where the non-vanishing components of the average velocity $\textbf{V}_s$ are given in Equations (\ref{vxi}) and (\ref{vyi}), and $n_{0,M}=n_0(1/2+{b})^2$ so that the density, current density and pressure ($P_{zz}$) calculated from the Maxwellian DF match with those calculated from the DF (\ref{quad_df2}). We see from Figures 3 and 4 that the deviation of the DFs from the Maxwellian DF (\ref{max_df}) is much larger than might be assumed from Figures 1 and 2. For increasing $u_{xi}/v_{th,i}$, there is an increase in the difference between the quadratic DF and the Maxwellian DF, which is to be expected, due to the appearance of multiple peaks in the DF for these parameters. We also note that, for the parameter values shown, the electron DFs are closer to Maxwellian than the ion DFs, due to the much smaller values of $u_{xe}/v_{th,e}$ for physically reasonable values of $T_e/T_i$.

Figures \ref{fig:quad_df_contour} ($z=0$) and \ref{fig:quad_df_contour2} ($z=0.5$) show contour plots of the ion DF in the $v_x$-$v_y$-plane (for $v_z=0$), for the same parameter values as before. The values for $v_y=0$ do not match up between the two figures, however, since the normalisation is different (here we have normalised so that the maximum value over the plane is one). Again, we can see the multiple peaks in the $v_x$-direction, this time for a range of $v_y$ values in each case, and note that for these parameters the DF is single-peaked in the $v_y$-direction. As before, the electron DFs will be single-peaked (and closer to Maxwellian) for these parameters due to the small values of $u_{xe}^2/v_{th,e}^2$.

Figures \ref{fig:quad_df_contour_diff} ($z=0$) and \ref{fig:quad_df_contour_diff2} ($z=0.5$) show contour plots of $f_i-f_{M,i}$ in the $v_x$-$v_y$-plane (for $v_z=0$), for the same parameter values as before. Again, we see that there is a substantial difference between the quadratic DF and the Maxwellian, even for small values of $u_{xi}/v_{th,i}$ when the DF looks close to Maxwellian. This difference is again seen to get larger as we increase $u_{xi}/v_{th,i}$.

We will now compare the plots of the quadratic DF with the examples given by \cite{Allanson-2015}, in which only single-peaked DFs are found. To do this, we define the magnetisation parameter, $\delta_s$, as
\begin{equation}
\delta_s=\frac{m_sv_{th,s}}{eB_0L},
\end{equation}
i.e. the ratio between the species gyroradius, $\rho_s=v_{th,s}/\Omega_s$ (for $\Omega_s=eB_0/m_s$ the species gyrofrequency) and the current sheet half-thickness, $L$. By using Equation (\ref{uxs_uys}), we can derive the following relation between $u_{xs}$ and $\delta_s$,
\begin{equation}
\frac{\vert u_{xs}\vert}{v_{th,s}}=2\delta_s\label{u delta}.
\end{equation} 
We can compare plots of the quadratic DF with the results of \cite{Allanson-2015} by using the same values of $\delta_s$ and $\beta_{pl}$ ($\delta_s=0.15$, $\beta_{pl}=0.85$). These are shown in Figures \ref{fig:compare1} (contour plots of $f_i$), \ref{fig:compare2} (contour plots of $f_i-f_{M,i}$), \ref{fig:compare3} (line plots of $f_i/f_{M,i}$ in the $v_x$-direction) and \ref{fig:compare4} (line plots of $f_i/f_{M,i}$ inthe $v_y$-direction). The $z$ values shown are $z=0$ and $z=1$, to match those chosen by \cite{Allanson-2015}. It is important to note that the Maxwellian DF used for comparison is normalised differently here than it is in the work by \cite{Allanson-2015} (so that the appropriate velocity moments are reproduced), but it is still useful to compare the departure from the respective Maxwellian DF in each case. The figures show that the quadratic DF (\ref{quad_df2}) is single-peaked for these parameter values, and that there is a significant departure from the Maxwellian DF, which has a different profile than that found by \cite{Allanson-2015}. In the $v_x$-direction, we see that the profile and values of $f_i/f_{M,i}$ for $z=0$ are not too different that that found by \cite{Allanson-2015}, but for $z=1$ the difference is much more noticeable. In the $v_y$-direction, the profiles of $f_i/f_{M,i}$ are qualitatively similar to those found by \cite{Allanson-2015}, but the values are significantly smaller in the quadratic case. 

Since, for the quadratic DF case ($N=2$), the lowest obtainable value of $\beta_{pl}$ is $1/2$, we cannot compare our results with those of \cite{Allanson-2016}, in which $\beta_{pl}=0.05$ is used (for our method, $N \ge 20$ would be necessary for a comparison). In that work, it was suggested that multiple maxima may occur for large values of $\delta_s/\beta_{pl}$, which corresponds to large $\delta_s$ here (due to the different definition of $u_{s}$ as discussed above). For such parameter ranges, however, numerical convergence could not be achieved for the Hermite polynomial sums, and so the behaviour of the DFs could not be investigated. Looking at the quadratic DF for such parameter ranges, we see that multiple maxima occur, i.e. as $u_{xs}$ ($\delta_s$) is increased. While this is not proof of the correctness of the conjecture made by Allanson et al. (2016), it is interesting that we now at least have an example for DFs exhibiting multiple maxima in the suggested parameter regime. 

The presence of multiple maxima in the DFs raises the question of whether they could be microscopically unstable. A full stability analysis is beyond the scope of this paper but, to get an indication of stability, we have applied the Penrose criterion (e.g. \cite{Krallbook-1973}) locally to the DFs, treating $z$ as a parameter (see Appendix \ref{sec:penrose}). Although this is not expected to provide a complete assessment of the stability properties of the DFs, it is reassuring that all the cases we investigated with $\beta_{pl}<1$ and multiple maxima were found to be stable according to the Penrose criterion.

\section{Summary and conclusions}
\label{sec:summary}

Up to now, with the exception of the linear force-free case \citep{Sestero-1967, Bobrova-2001}, only one single example of an explicitly known DF for collisionless force-free current sheets allowing plasma beta values smaller than one was known \citep{Allanson-2015,Allanson-2016}.  This DF can only be expressed as an infinite sum of Hermite functions with very slow convergence, which makes it cumbersome to use in practice. In this paper, we have presented a method that allows us to find further equilibrium DFs for force-free current sheets with a plasma beta smaller than one, but which consist of a finite number of terms and are, therefore, easier to use. As in the work by \cite{Allanson-2015, Allanson-2016}, the method is based on a transformation of the pressure, as suggested by \cite{Harrison-2009a}. The transformation used in this paper is an integer power $N$ of the original pressure, allowing in principle for a reduction of the plasma beta by a factor of $1/N$. 

%


For the example of the force-free Harris sheet, we have derived the corresponding DFs for each $N$ in closed form and, to provide an illustrative example, have given a detailed analysis of 
the quadratic case ($N=2$), giving a much more  comprehensive discussion than that previously given by \cite{Neukirch-2017}. An interesting aspect of this detailed analysis is that the
%
%
conditions on the parameters for the quadratic case are much less restrictive  than assumed by \cite{Neukirch-2017}. These less restrictive conditions allow us to achieve multi-peaked DFs that give rise to a plasma beta of lower than one, whereas, for the parameter range considered by \cite{Neukirch-2017}, only single-peaked DFs were found. In the work by \cite{Allanson-2015,Allanson-2016}, only single-peaked DFs were found, which is an obvious difference between the two models. It must be emphasised, however, that in the work by \cite{Allanson-2015,Allanson-2016}, the full parameter range could not be explored since numerical convergence of the infinite sums could not always be achieved. For the parameters used by \cite{Allanson-2015}, the new DFs presented in this paper are single-peaked. 

In the quadratic case example, the plasma beta is bounded from below by $1/2$, which is a disadvantage compared with the DFs found 
by \cite{Allanson-2015,Allanson-2016}, in which the plasma beta could be arbitrarily small. However, the minimum value of the plasma beta can in principle be made smaller by 
making the power $N$ in the transformation larger.
As demonstrated in this paper, for the general $N$ case, it is possible to derive both DFs and conditions on the parameters that ensure strict neutrality and consistency between the macroscopic and microscopic parameters of the equilibrium. The conditions for the positivity of the DF will have to be considered along the same lines as presented in this paper for the quadratic case, but on a case-by-case basis. \\

The authors would like to thank the anonymous referees, whose comments have helped to improve this manuscript. We acknowledge the support of the Science and Technology Facilities Council via the consolidated grants ST/K000950/1 and ST/N000609/1 and the doctoral training grant ST/K502327/1 (O. A.), and the Natural Environment Research Council via grant no. NE/P017274/1 (Rad-Sat) (O. A.). F. W. and T. N. would also like to thank the University of St Andrews for general financial support.

\appendix

\section{The ``full" plasma beta for the quadratic case}
\label{sec:full_beta}

As stated in Section \ref{sec:pressure_transformation}, throughout this paper we have defined the plasma beta as $\beta_{pl}=P_{zz}/(B^2/(2\mu_0))$, for $P_{zz}$ the $zz$-component of the pressure tensor. It may also be of interest, however, to investigate the effect of the pressure transformation on the quantity $\bar\beta_{pl}=P/(B^2/(2\mu_0))$, where $P=(P_{xx}+P_{yy}+P_{zz})/3$, since this is the more conventional definition of the plasma beta. We will illustrate this for the quadratic pressure transformation. Using the definitions of the diagonal pressure tensor components, $P$ can be calculated as
\begin{eqnarray}
P=\frac{1}{3}\sum_sm_s\left(\int \textbf{v}^2f_s\mathrm{d}^3v-n_s\textbf{V}_s^2\right).\label{TrP}
\end{eqnarray}
Using Equations (\ref{B0}), (\ref{quad_ns}), (\ref{vxi}) and (\ref{vyi}), and substituting the DF (\ref{quad_df2}) into Equation (\ref{TrP}) gives, for the quadratic case,
\begin{eqnarray}
\bar\beta_{pl}&=&\frac{1}{2}\left(\frac{1}{2}+b\right)+\frac{1}{3}\frac{\beta_e}{\beta_e+\beta_i}\left(1+\frac{T_e^2}{T_i^2}\frac{m_e}{m_i}\right)\frac{u_{xi}^2}{v_{th,i}^2}\left(\frac{4b\mbox{sech}^2(z/L)}{1+2b}-2\right).
\end{eqnarray}
An equivalent expression can be derived for Harrison and Neukirch's DF for the untransformed pressure (\cite{Harrison-2009b, Neukirch-2009}) as
\begin{eqnarray}
\bar\beta_{pl, H\&N}&=&\left(\frac{1}{2}+b\right)+\frac{1}{3}\frac{\beta_e}{\beta_e+\beta_i}\left(1+\frac{T_e^2}{T_i^2}\frac{m_e}{m_i}\right)\frac{u_{xi}^2}{v_{th,i}^2}\left(\frac{4b\mbox{sech}^2(z/L)}{1+2b}-\frac{1}{2}\right).
\end{eqnarray}
Clearly, the quadratic pressure transformation has also had the effect of reducing $\bar\beta_{pl}$. In the limit $u_{xi}\to0$, we have $\bar\beta_{pl}=\beta_{pl}$ in each case.

\section{The plasma beta for summative pressure functions}
\label{sec:summative}

It can be seen from various current sheet models discussed in the literature (e.g. \cite{Channell-1976, Attico-1999,Harrison-2009b,Neukirch-2009, Wilson-2011,Abraham-Shrauner-2013,Kolotkov-2015,Wilson-2017}) that, if the pressure function is of a summative form in terms of its dependence on the vector potential components, i.e. 
\begin{equation}
P(A_x,A_y)=P_1(A_x)+P_2(A_y),\label{p sum}
\end{equation}
 then the plasma beta is constrained to be at least one. This has been considered in a more general sense by \cite{Allanson-thesis}; in this appendix, we will summarise the discussion given therein.

Firstly, for summative pressure functions of the form in Equation (\ref{p sum}), the force balance equation for force-free fields, given by
\begin{equation}
P_{zz}(A_x,A_y)+\frac{B_{0}^2}{2\mu_0}=P_T,
\end{equation}
can be split into the two equations
\begin{eqnarray}
P_1(A_x)+\frac{1}{2\mu_0}B_y^2(A_x)=P_{T1},\nonumber\\
P_2(A_y)+\frac{1}{2\mu_0}B_x^2(A_y)=P_{T2},\label{force_balance}
\end{eqnarray}
where $P_{T1}+P_{T2}=P_T$ is the total pressure (for constants $P_{T1}$ and $P_{T2}$).

Secondly, as discussed by \cite{Bobrova-2001} and \cite{Vekstein-2002}, all 1D force-free magnetic fields can be written in the general form
\begin{equation}
\textbf{B}(z)=B_0(\cos\left(S(z)\right), \sin\left(S(z)\right), 0),\label{general_ff}
\end{equation}
where $S(z)=\int\alpha(z)\mathrm{d}z$, with $\alpha$ the force-free parameter as discussed in Section \ref{sec:intro}. For linear force-free fields (\cite{Channell-1976, Sestero-1967, Attico-1999, Bobrova-2001}), the function $S(z)$ must be a linear function of $z$, whereas for nonlinear force-free fields the function $S(z)$ can, in principle, be any differentiable function.

We will now assume that
\begin{enumerate}
\item $P_{1}(A_x)\ge0$ and $P_2(A_y)\ge0$,
\item there exist points $z=z_1$, $z=z_2$ such that 
\begin{equation}
\sin^2(S(z_1))=1, \mbox{  }\cos^2(S(z_1))=0,  \mbox{  }\sin^2(S(z_2))=0, \mbox{  } \cos^2(S(z_2))=1.
\end{equation}
\end{enumerate}

Assumption (i) can be justified by considering the inverse problem defined by Equation (\ref{inverse_prob}), in which the dependence of $P_{zz}$ on $A_x$ and $A_y$ is tied to the dependence of the DF of $p_{xs}$ and $p_{ys}$, respectively. Since the DF must be positive with respect to the independent variation of $p_{xs}$ and $p_{ys}$, it follows that $P_{zz}$ must be positive with respect to independent variations of $A_x$ and $A_y$.

Assumption (ii) is always true for a linear force-free field, since $S(z)$ is a linear function. It will also hold for a particular class of nonlinear force-free fields, in which one of the magnetic field components goes through zero, and the other tends to zero at $\pm\infty$. The force-free Harris sheet (Equation (\ref{Bffhs})) is an example of such a magnetic field profile.
Combining the assumptions made above, we have that
\begin{eqnarray}
P_{T1}&=&P_1(A_x(z_1))+\frac{B_0^2}{2\mu_0}\sin^2(S(z_1))\ge\frac{B_0^2}{2\mu_0}\label{ineq1},
\nonumber\\
P_{T2}&=&P_2(A_y(z_2))+\frac{B_0^2}{2\mu_0}\cos^2(S(z_2))\ge\frac{B_0^2}{2\mu_0}\label{ineq2},
\end{eqnarray}
for the particular points $z_1$ and $z_2$ defined above. However, since $P_{zz}$ is constant, and $P_{T1}$ and $P_{T2}$ are independent of each other through separation of variables, the conditions in Equations (\ref{ineq1}) must be valid for all values of $z$. This fact, together with the Equations (\ref{force_balance}), gives
\begin{equation}
P_T=P_{T1}+P_{T2}\ge2\frac{B_0^2}{2\mu_0}\Rightarrow P_1(A_x)+P_2(A_y)+\frac{B_0^2}{2\mu_0}\ge2\frac{B_0^2}{2\mu_0}.
\end{equation}
Finally, dividing through by the magnetic pressure, ${B_0^2}/(2\mu_0)$, gives
\begin{equation}
\beta_{pl}+1\ge2\Rightarrow \beta_{pl}\ge1.
\end{equation}
This discussion demonstrates that, for linear force-free fields and certain types of nonlinear force-free fields, the plasma beta is constrained to be at least one whenever the pressure function is assumed to have a summative form.

\section{Distribution functions for a general $N$}
\label{sec:general dfs}

The pressure function (\ref{pzz_macro}) can be written in the expanded form
\begin{eqnarray}
\bar{P}_{zz}&=&\frac{B_0^2}{2\mu_0}\frac{(1/2+{b})^{1-N}}{N}\sum_{k=0}^{N}\sum_{l=0}^{N-k}\binom{N}{k}\binom{N-k}{l}\frac{{b}^l}{2^{N-k-l}}\cos^{N-k-l}\left(\frac{2A_x}{B_0L}\right)\exp\left(\frac{2kA_y}{B_0L}\right).\nonumber\\\label{pzz_macro_expanded}
\end{eqnarray} 

The function (\ref{pzz_macro_expanded}) can be substituted into Equation (\ref{inverse_prob}), which can then be solved (e.g. by using Fourier/Weierstrass transforms) to give the DF as
\begin{eqnarray}
f_s(H_s,p_{xs}, p_{ys})&=&f_{0s}e^{-\beta_sH_s}\sum_{k=0}^{N}\sum_{l=0}^{N-k}\sum_{m=0}^{N-k-l}\binom{N}{k}\binom{N-k}{l}\binom{N-k-l}{m}\frac{{b}^l}{2^{2(N-k-l)}} \nonumber\\
&{}&\times e^{\left[(2m-(N-k-l))^2{u_{xs}^2}-k^2u_{ys}^2\right]/{2v_{th,s}^2}} e^{-i(2m-(N-k-l))\beta_su_{xs}p_{xs}}e^{k\beta_su_{ys} p_{ys}},\nonumber\\
\end{eqnarray}
where we have introduced the constant microscopic parameters $f_{0s}$, $u_{xs}$ and $u_{ys}$. The moments of this DF can be calculated as demonstrated in Section \ref{sec:quadratic} for the $N=2$ case, giving conditions between the microscopic and macroscopic parameters of the equilibrium.

For a physically meaningful solution, we need to ensure that $f_s\ge0$; this can be satisfied for $N=2$, as is shown in Appendix \ref{app:gs+}. For $N>2$, one would proceed along the same lines as for $N=2$, but the calculation becomes more involved.

\section{Positivity of the quadratic distribution function}
\label{app:gs+}

In this appendix, we will show that a necessary and sufficient condition for positivity of the quadratic DF (\ref{quad_df2}) over the whole phase space is
\begin{equation}
{b}\ge\frac{e^{\bar{u}_{xs}^2/2}}{2\sqrt{2}}\left(e^{\bar{u}_{xs}^2}+1\right)^{1/2}.\label{pos_condition}
\end{equation}
Positivity of the DF requires positivity of the function
\begin{eqnarray}
g_s(X,Y)&=&\frac{1}{8}e^{2\alpha}\cos(2X)+e^{-2\alpha}e^{2Y}+e^Y\cos{X}+{b}e^{\alpha/2}\cos{X}+2{b}e^{-\alpha/2}e^Y+{b}^2+\frac{1}{8},\nonumber\\
&{}&
\end{eqnarray}
over the whole phase space, where $X=\beta_su_{xs}p_{xs}$, $Y=\beta_su_{ys}p_{ys}$ and $\alpha=\bar{u}_{xs}^2$ (note that $X$, $Y$ and $\alpha$ are species dependent but we do not show $s$ subscripts for notational convenience). Using that fact that $\cos(2X)=2\cos^2X-1$ and writing
\begin{eqnarray}
\frac{1}{4}e^{2\alpha}\cos^2X+e^Y\cos{X}=\frac{1}{4}e^{2\alpha}\left(\cos{X}+2e^{-2\alpha}e^Y\right)^2-e^{-2\alpha}e^{2Y},
\end{eqnarray}
we see that $g_s\ge0$ requires
\begin{equation}
\frac{1}{4}e^{2\alpha}\left(\cos{X}+2e^{-2\alpha}e^{Y}\right)^2+\left(2e^{-\alpha/2}e^{Y}+e^{\alpha/2}\cos{X}\right){b}+{b}^2+\frac{1}{8}(1-e^{2\alpha})\ge0.\label{gs+}
\end{equation}

\subsection{$p_{ys}$-direction}
We first consider the $p_{ys}$-direction (i.e. the $Y$-direction) and look to minimise $g_s$ with respect to $\xi=e^Y$. We have
\begin{equation}
\frac{\partial g_s}{\partial\xi}=2e^{-2\alpha}\xi+\cos X +2{b}e^{-\alpha/2},
\end{equation}
which vanishes for
\begin{equation}
\xi=\xi_{\mbox{min}}=-\frac{e^{2\alpha}}{2}\left(\cos{X} +2{b}e^{-\alpha/2}\right).
\end{equation}
i.e. when $Y=Y_{\mbox{min}}=\ln\left\vert\xi_{\mbox{min}}\right\vert$. This gives
\begin{eqnarray}
g_s(X,Y_{\mbox{min}})=-\left(e^\alpha-1\right){b}^2-e^{\alpha/2}\cos{X}\left(e^\alpha-1\right){b}-\frac{1}{8}\left(e^{2\alpha}-1\right),\label{gs_ymin}
\end{eqnarray}
which is a minimum since
\begin{equation}
\frac{\partial^2 g_s}{\partial\xi^2}\Bigg\vert_{\xi=\xi_{\mbox{\tiny{min}}}}=2e^{-2\alpha}>0.
\end{equation}
We require $g_s(X,Y_{\mbox{min}})>0$. If we assume that $\xi_{\mbox{min}}>0$, i.e that ${b}<-e^{\alpha/2}/2$, we see that $g_s(X,Y_{\mbox{min}})>0$ cannot be satisfied for all $X$; the first and third terms on the right-hand side (RHS) of Equation (\ref{gs_ymin}) are always negative (because $\alpha\ge0$). The second term can be positive or negative for some range of $X$ values, and so the RHS cannot be positive for all $X$. Note also that ${b}<-1/2$ would lead to a negative pressure through Equation (\ref{pff_quad}), and so it would not make sense if this parameter range were found to give rise to positive DFs over the whole phase space. We assume, therefore, that $\xi_{\mbox{min}}\le0$, which gives ${b}\ge e^{\alpha/2}/2$. However, $\xi$ must be at least zero, since $\xi=e^Y$. Therefore, since $g_s$ is an increasing function of $\xi$ in this range (a quadratic), the condition for positivity of $g_s$ is given by
\begin{equation}
\lim_{Y\to-\infty}g_s(X,Y)\ge0,
\end{equation}
since $Y\to-\infty$ gives $e^Y=0$. 

\subsection{$p_{xs}$-direction}
We now consider the $p_{xs}$-direction (i.e. the $X$-direction). We wish to minimise the function
\begin{equation}
G_0(X)=\lim_{Y\to-\infty}g_s(X,Y).
\end{equation}
The first and second derivatives of $G_0$ are given by
\begin{eqnarray}
\frac{\mathrm{d}G_0}{\mathrm{d}X}&=&-\left(\frac{1}{2}e^{2\alpha}\cos{X}+{b}e^{\alpha/2}\right)\sin{X},\nonumber\\
\frac{\mathrm{d}^2G_0}{\mathrm{d}X^2}&=&-\frac{1}{2}e^{2\alpha}\cos{2X}-{b}e^{\alpha/2}\cos{X}.
\end{eqnarray}
and so maxima/minima of $G_0$ occur when (a) $\cos{X}=-2{b}e^{-3\alpha/2}$ and (b) $X=p\pi$ (for integer $p$). In case (a), the second derivative of $G_0$ equals $e^{2\alpha}/2-2{b}^2e^{-\alpha}$, and so this case will give a minimum of $G_0$ when ${b}\le e^{3\alpha/2}/2$. If this is true, we have $\cos{X_{\mbox{min}}}=-2{b}e^{-3\alpha/2}$, and so 
\begin{equation}
G_0(X_{\mbox{min}})={b}^2(1-e^{-\alpha})+\frac{1}{8}(1-e^{2\alpha}),
\end{equation}
which will be positive if
\begin{equation}
{b}^2\ge\frac{e^\alpha}{8}(e^\alpha+1),\label{pos1}
\end{equation}
provided ${b}\le e^{3\alpha/2}/2$. Since $\alpha\ge0$, Equation (\ref{pos1}) gives either ${b}\ge1/2$ or ${b}\le-1/2$ for $G_0\ge0$. We have already discounted the case ${b}\le-1/2$, however, since this would give $\xi_{\mbox{min}}>0$, for which we cannot have $g_s>0$ for all $X$. The condition (\ref{pos1}) can, therefore, be expressed as
\begin{equation}
\frac{e^{\alpha/2}}{2\sqrt{2}}\left(e^{\alpha}+1\right)^{1/2}\le{b}\le e^{3\alpha/2}/2.
\end{equation}

We now consider case (b), when $X=p\pi$ (for integer $p$). In this case, the second derivative of $G_0$ is equal to $-e^{2\alpha}/2-{b}e^{\alpha/2}(-1)^p$ (since $\cos(p\pi)=(-1)^p$). When $p$ is even, we arrive at the condition ${b}<-e^{3\alpha/2}/2$ for a minimum of $G_0$. This is not allowed, however, since it can give $\xi_{\mbox{min}}>0$, for which we cannot have $g_s>0$ for all $X$, as previously discussed. We assume, therefore, that $p=2q+1$ for integer $q$, i.e. that it is odd, in which case we require ${b}>e^{3\alpha/2}$ for a minimum of $G_0$. For $X=(2q+1)\pi$, we have
\begin{equation}
G_0((2q+1)\pi)={b}^2-{b}e^{\alpha/2}+\frac{1}{8}(1+e^{2\alpha}).\label{b_quad}
\end{equation}
The minimum of the RHS of Equation (\ref{b_quad}) (with respect to ${b}$) is zero and so, when $X=(2q+1)\pi$, the condition $G_0\ge0$ is always satisfied.

Based on the analysis above, we conclude that positivity of the DF (\ref{quad_df2}) can be ensured when the condition (\ref{pos_condition}) is satisfied. Since $\alpha\ge0$, we always have 
\begin{equation}
\frac{e^{\bar{u}_{xs}^2/2}}{2\sqrt{2}}\left(e^{\bar{u}_{xs}^2}+1\right)^{1/2}\le e^{3\alpha/2}/2,
\end{equation}
so that the lower bound for ${b}$ to have $g_s\ge0$ is the expression on the left-hand side.

\section{Stability - the Penrose criterion}

\label{sec:penrose}

The Penrose criterion (see, e.g., \cite{Krallbook-1973}) states that an equilibrium DF with a local minimum occurring at $w=w_0$ is potentially unstable to the growth of electrostatic waves if 
\begin{equation}
P(F) = \int_{-\infty}^{\infty}\frac{F(w_0)-F(w)}{(w-w_0)^2}\mathrm{d}w<0,\label{PofF}
\end{equation}
with
\begin{equation}
F(w) = F_e(w)+\frac{m_e}{m_i}F_i(w),\label{penrose_F}
\end{equation}
where $F_s$ is defined in terms of the equilibrium DF for species $s$ as
\begin{equation}
F_s(w) = \int\delta\left(w-\frac{\textbf{k}\boldsymbol{\cdot}\textbf{v}}{k}\right)f_s\mathrm{d}^3v,
\end{equation}
where $k=|\textbf{k}|$ ($\textbf{k}$ is the wave vector). For the DF (\ref{quad_df2}), the function $F_s$ is given by
\begin{eqnarray}
F_{s}(\bar{w}_s) &=& \frac{n_0e^{-\bar{w}_s^2/2}}{\sqrt{2\pi}v_{th,s}}\Bigg\{\frac{1}{8}e^{2\bar{u}_{xs}^2(1-S_1)}\cos(2S_2\bar{u}_{xs}\bar{w}_s+2T)\nonumber\\
&{}&+F_{1s}e^{2F_{2s}\bar{w}_s}+F_{3s}e^{F_{2s}\bar{w}_s}\cos\left(S_2\bar{u}_{xs}\bar{w}_s+F_{4s}\right)+(b^2+1/8)\nonumber\\
&{}&+be^{\bar{u}_{xs}^2(1-S_1)/2}\cos\left(S_2\bar{u}_{xs}\bar{w}_s+T\right)
+F_{5s}e^{F_{2s}\bar{w}_s}
\Bigg\},\label{Fi}
\end{eqnarray}

with $\bar{w}_s=w/v_{th,s}$, $\bar{u}_{xs}=u_{xs}/v_{th,s}$, $\bar{u}_{ys}=u_{ys}/v_{th,s}$ and 
\begin{eqnarray}
S_1&=&1-\cos^2\phi\sin^2\theta,\\
S_2&=&\sin\theta\cos\phi,\\
T&=&4\tan^{-1}\left(e^{z/L}\right),\\
F_{1s}&=&\mbox{sech}^4(z/L)\exp\left(-2\bar{u}_{ys}^2\sin^2\theta\sin^2\phi\right),\\
F_{2s}&=&\bar{u}_{ys}\sin\theta\sin\phi,\\
F_{3s}&=&\mbox{sech}^2(z/L)\exp\left(-\frac{\bar{u}_{ys}^2}{2}\left(1-\cos^2\theta-2\cos^2\phi\sin^2\theta\right)\right),\\
F_{4s}&=&T-\bar{u}_{xs}\bar{u}_{ys}\sin^2\theta\cos\phi\sin\phi,\\
F_{5s}&=&2b\mbox{sech}^2(z/L)\exp\left(-\frac{\bar{u}_{ys}^2}{2}\sin^2\theta\sin^2\phi\right),
\end{eqnarray}
where $\theta$ and $\phi$ are the inclination and azimuthal angles of the wave vector, respectively.

We have investigated properties of the function $F(w)$ for various parameter values and angles of propagation. To illustrate our findings, we will focus on the case with $(u_{xi}/v_{th,i})^2=1.2$ and $\beta_{pl}=0.95$, as shown in Figure \ref{fig:quad_df} (f) (for example). We do this since, out of the different cases we considered in Section \ref{sec:df_plots}, this was the case with the most pronounced double maxima in the $v_x$-direction. We mentioned previously that the double maxima in the DF can become even more pronounced for higher values of the plasma beta, but we will not consider these cases since the focus of this paper is on DFs for which $\beta_{pl}<1$.

As discussed in Section \ref{sec:df_plots}, the ion DFs, $f_i$, given by Equation (\ref{quad_df2}), can have multiple maxima in $v_x$, whereas the electron DFs have only one maximum due to the parameter restrictions. The function $F_i$, given by Equation (\ref{Fi}), can also have double maxima, whereas $F_e$ has only one maximum. Since $v_{th,e}={(m_i/m_e)^{1/2}(T_e/T_i)^{1/2}}v_{th,i}$, $F_e$ will typically be much wider than $F_i$. An example of the resulting structure of $F$ is shown in Figure \ref{fig:penrose}, where we have chosen the parameters $\beta_{pl}=0.95$, $(u_{xi}/v_{th,i})^2=1.2$, $z=0.3$, $T_i/T_e=1$, $\theta=\pi/2$ and $\phi=0$ (i.e. propagation in the $x$-direction - the direction in which $f_i$ has double maxima). From Figure \ref{fig:penrose} (a), we see that there is a double maximum from $F_i$ but, from Figures \ref{fig:penrose} (b), (c) and (d), we see that this double maximum is fairly insignificant due to the much wider $F_e$.  For the parameter values from Figure \ref{fig:penrose}, we calculated the Penrose function (\ref{PofF}) as $P(F)\approx0.0059$. If we take a higher temperature ratio, $F_e$ becomes less wide in comparison to $F_i$, but the resulting full $F$ is then less likely to have a double maximum at all. This example gives a good illustration of what appears happen for the other parameter sets we have tried. In all these cases, the Penrose function $P(F)$ is positive. For parameter sets with $\beta_{pl}<1$, therefore, we expect that our DFs are stable according the the Penrose criterion.



\clearpage
\setcounter{figure}{0} \renewcommand{\thefigure}{\arabic{figure}}

\begin{figure}
\centering\
\subfigure[]{\scalebox{0.32}{\includegraphics{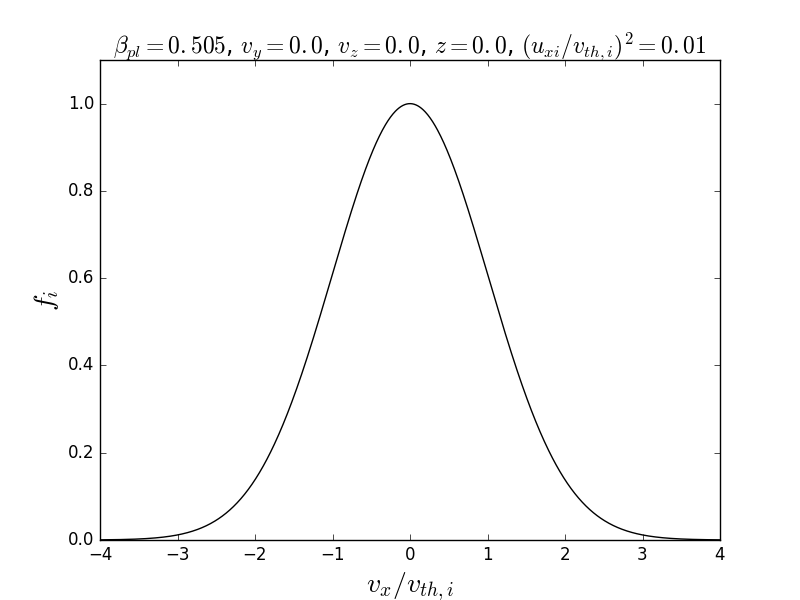}}}
\subfigure[]{\scalebox{0.32}{\includegraphics{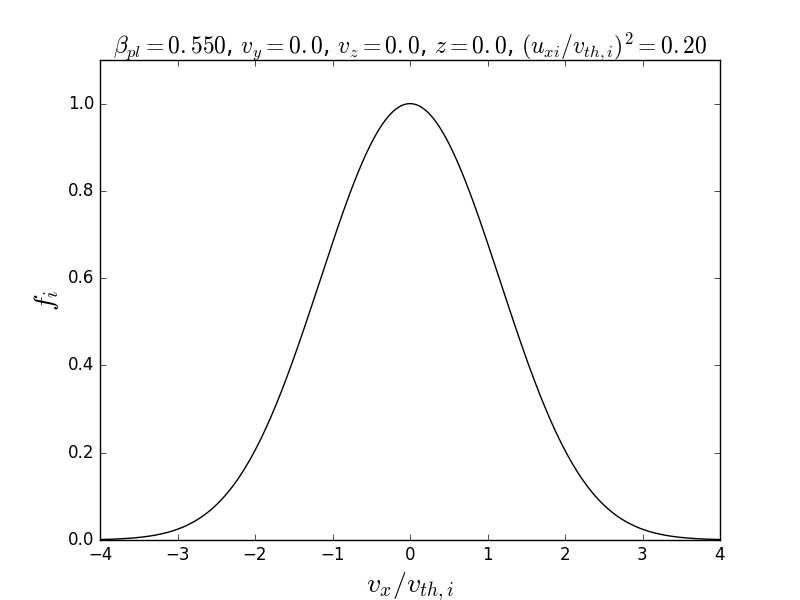}}}
\subfigure[]{\scalebox{0.32}{\includegraphics{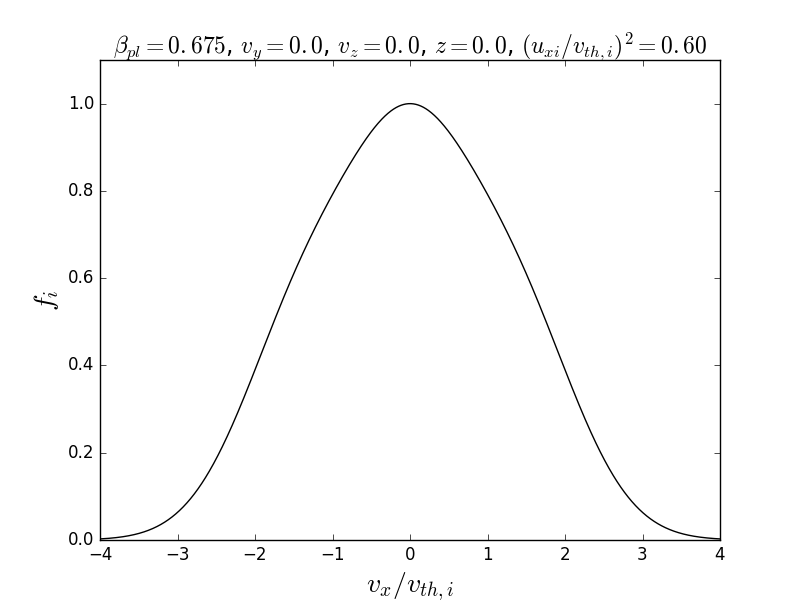}}}
\subfigure[]{\scalebox{0.32}{\includegraphics{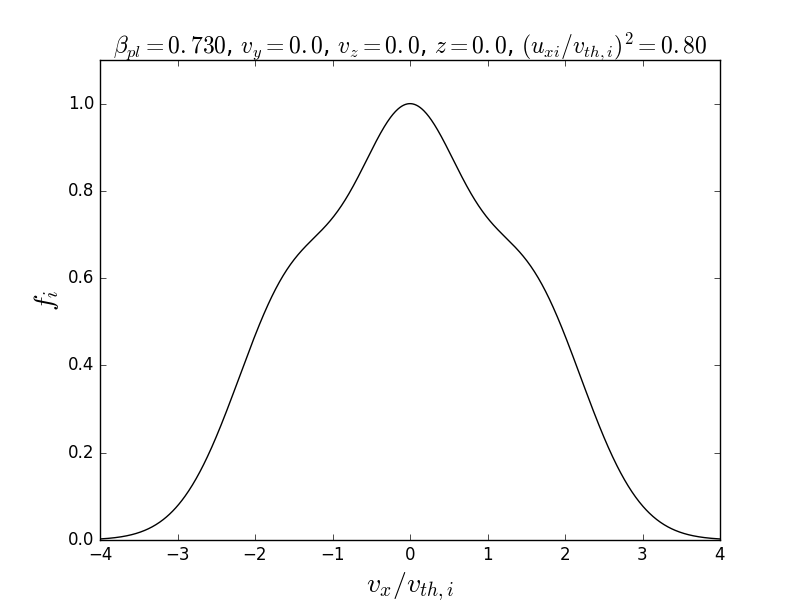}}}
\subfigure[]{\scalebox{0.32}{\includegraphics{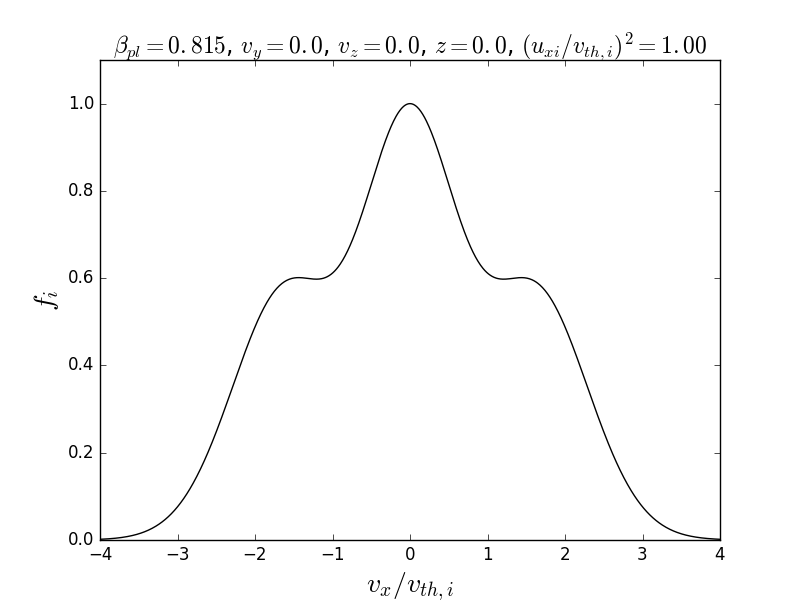}}}
\subfigure[]{\scalebox{0.32}{\includegraphics{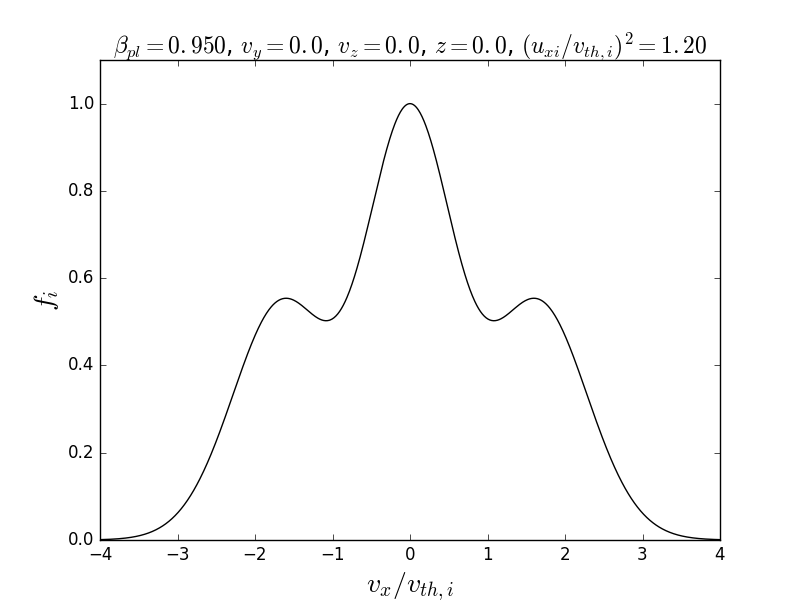}}}
\caption{Plots of the ion DF (\ref{quad_df2}) in the $v_x$-direction (with $v_y=v_z=z=0$) for various values of $u_{xi}/v_{th,i}$ and, hence, the plasma beta. The DFs are normalised to have a maximum value of one in each case.}
\label{fig:quad_df}
\end{figure}
\begin{figure}
\centering\
\subfigure[]{\scalebox{0.32}{\includegraphics{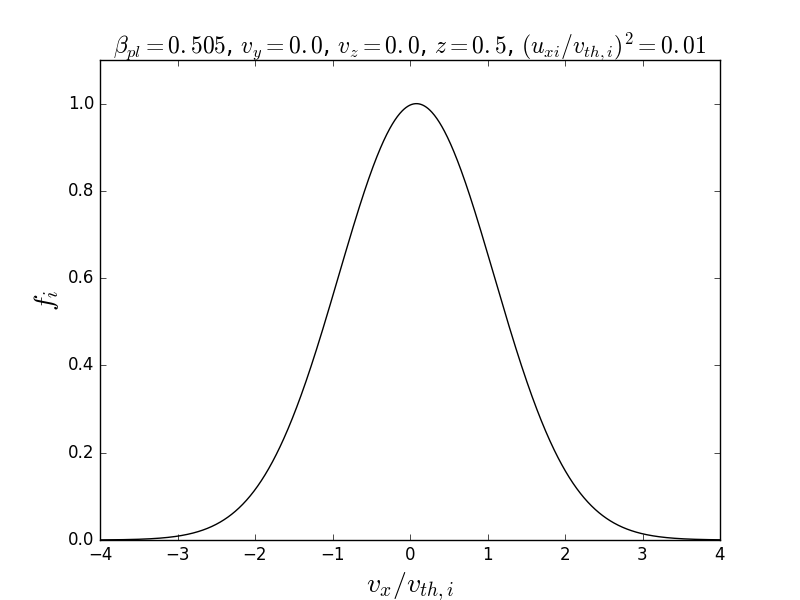}}}
\subfigure[]{\scalebox{0.32}{\includegraphics{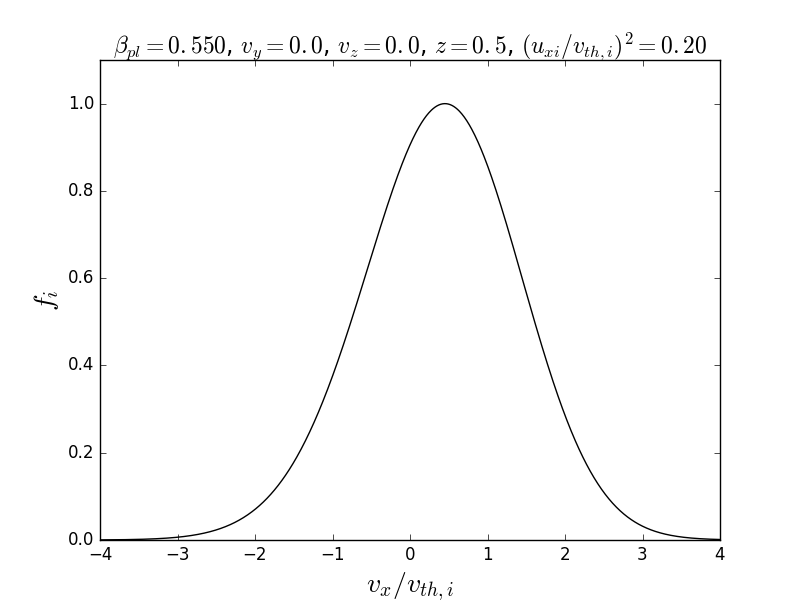}}}
\subfigure[]{\scalebox{0.32}{\includegraphics{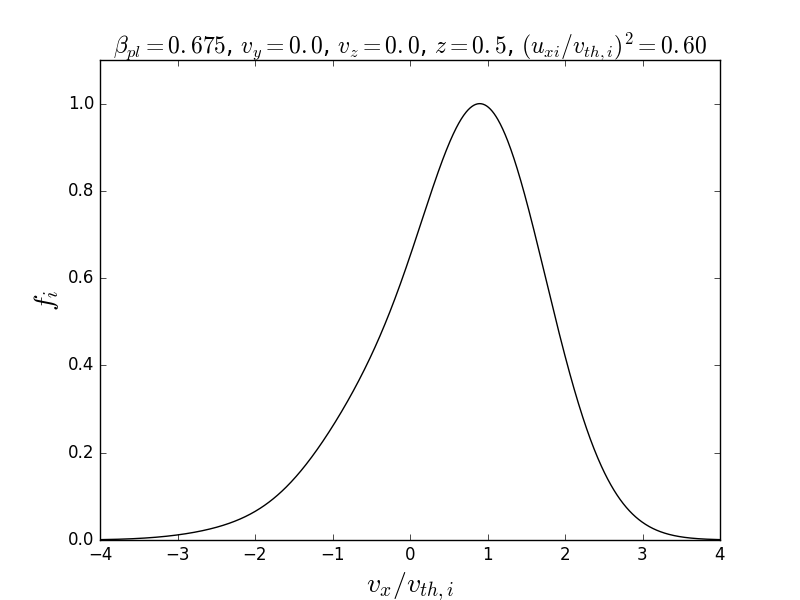}}}
\subfigure[]{\scalebox{0.32}{\includegraphics{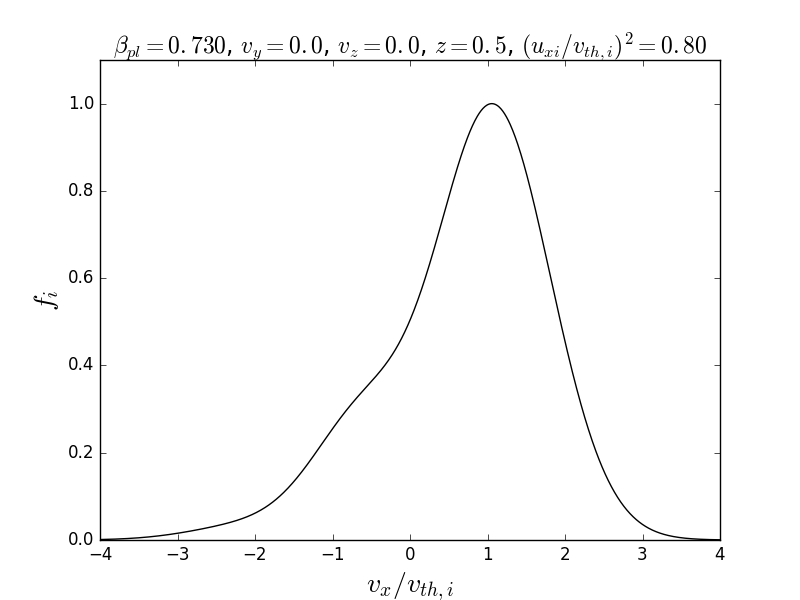}}}
\subfigure[]{\scalebox{0.32}{\includegraphics{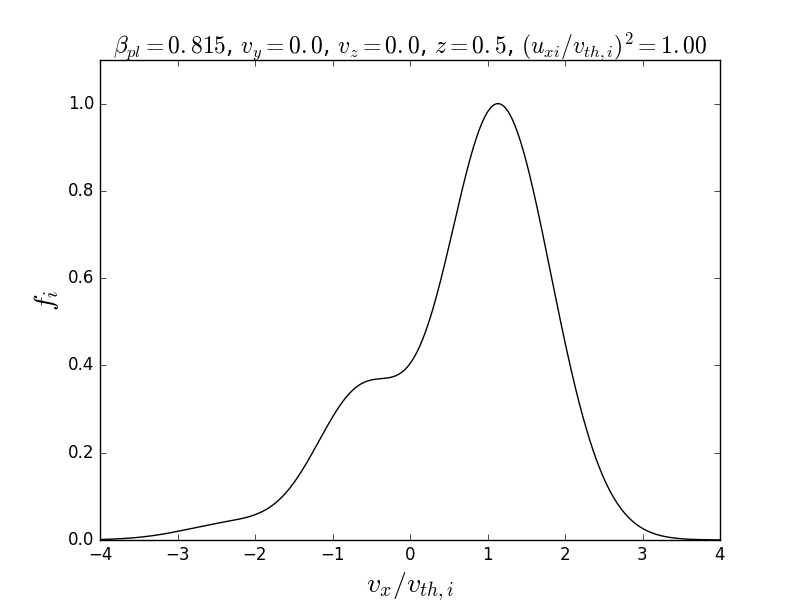}}}
\subfigure[]{\scalebox{0.32}{\includegraphics{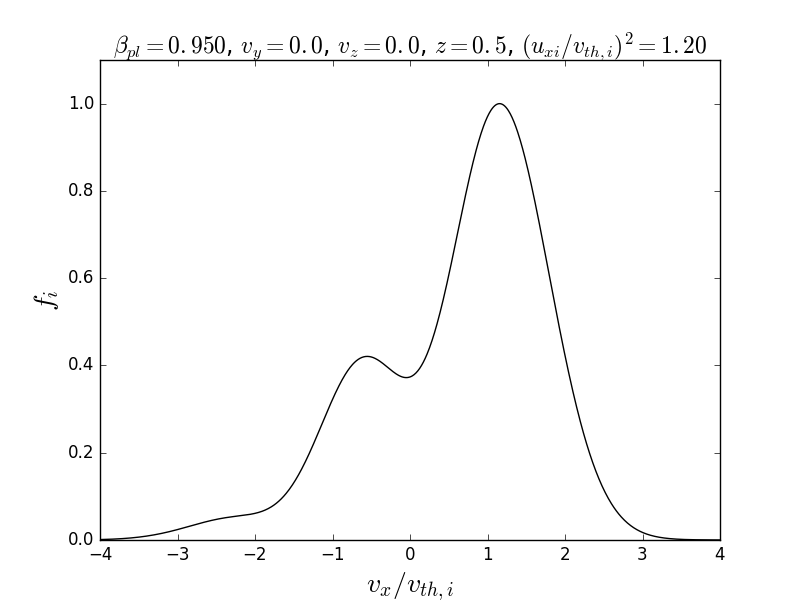}}}
\caption{Plots of the ion DF (\ref{quad_df2}) in the $v_x$-direction (with $v_y=v_z$ and $z=0.5$) for various values of $u_{xi}/v_{th,i}$ and, hence, the plasma beta. The DFs are normalised to have a maximum value of one in each case.}
\label{fig:quad_df2}
\end{figure}

\begin{figure}
\centering\
\subfigure[]{\scalebox{0.32}{\includegraphics{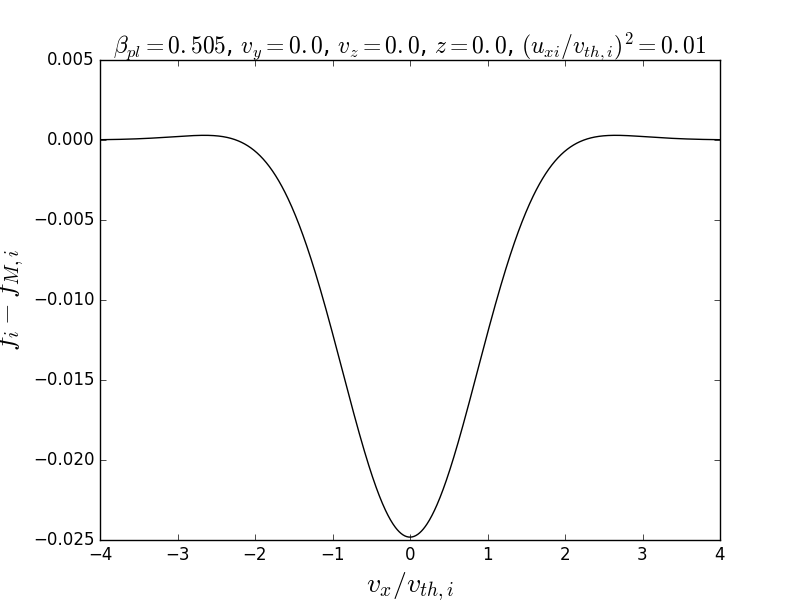}}}
\subfigure[]{\scalebox{0.32}{\includegraphics{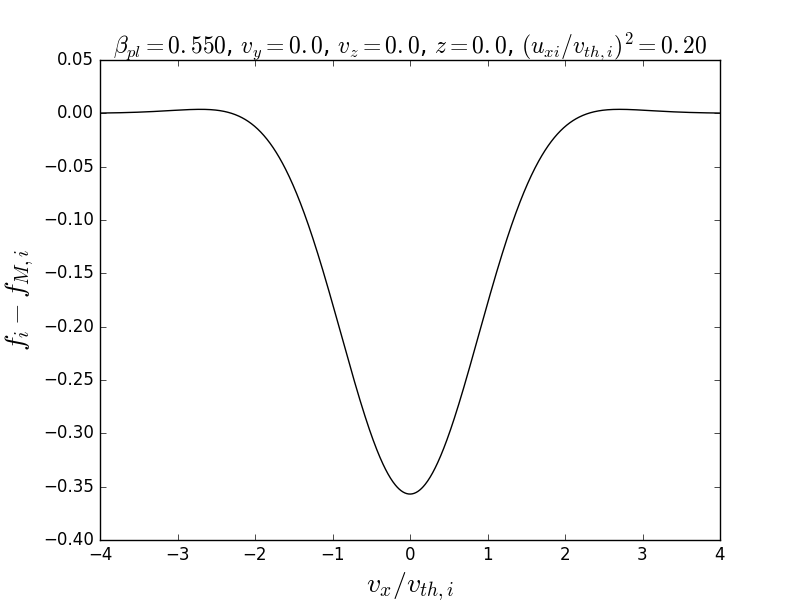}}}
\subfigure[]{\scalebox{0.32}{\includegraphics{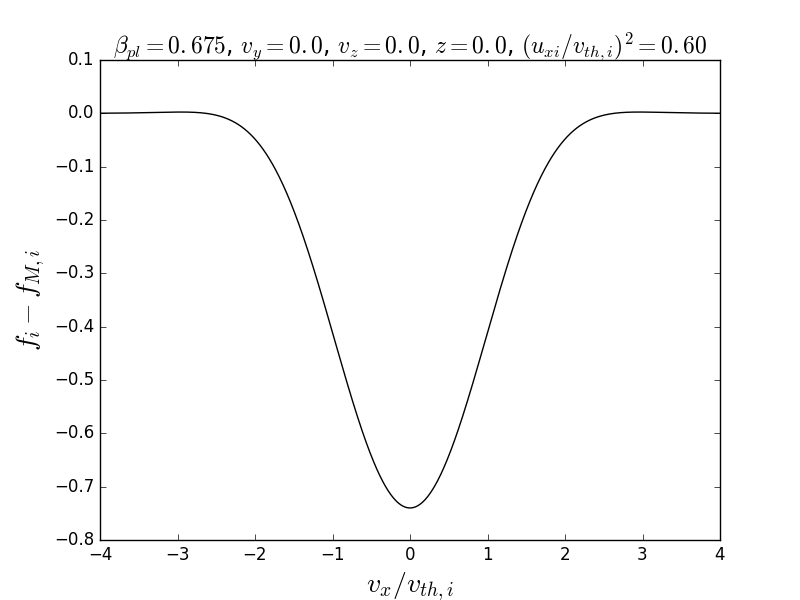}}}
\subfigure[]{\scalebox{0.32}{\includegraphics{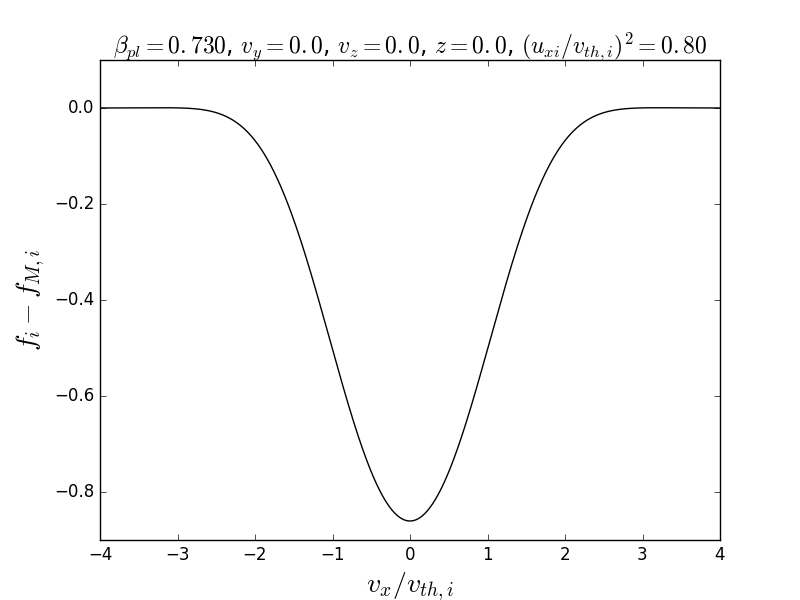}}}
\subfigure[]{\scalebox{0.32}{\includegraphics{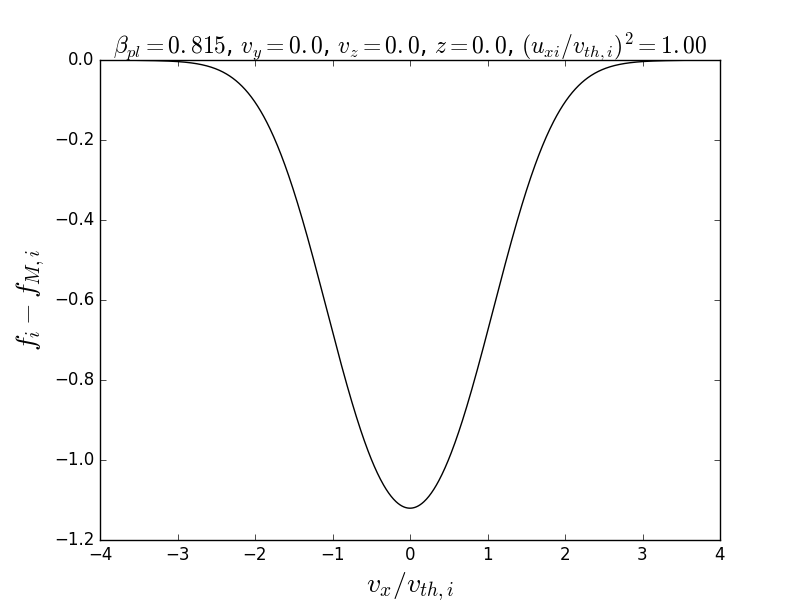}}}
\subfigure[]{\scalebox{0.32}{\includegraphics{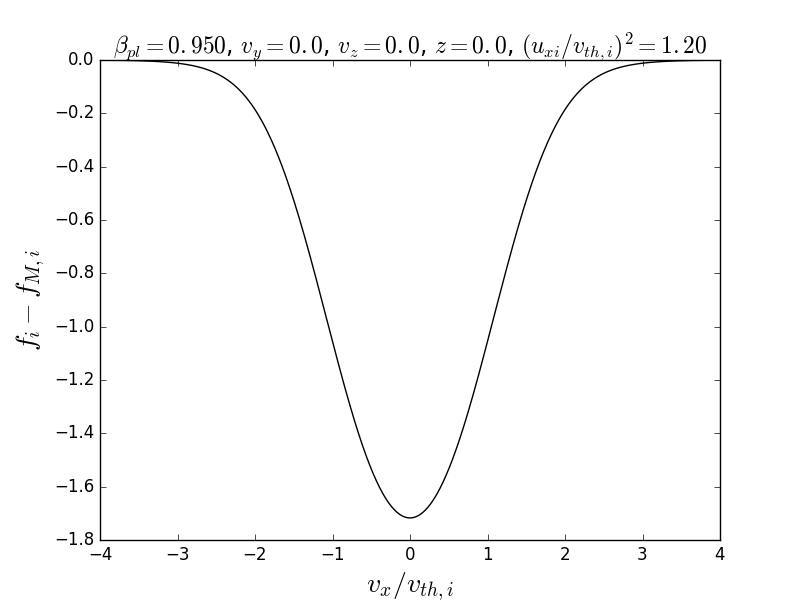}}}
\caption{Plots of $f_i-f_{M,i}$ in the $v_x$-direction (with $v_y=v_z=z=0$) for various values of $u_{xi}/v_{th,i}$ and, hence, the plasma beta.}
\label{fig:quad_df_diff}
\end{figure}

\begin{figure}
\centering\
\subfigure[]{\scalebox{0.32}{\includegraphics{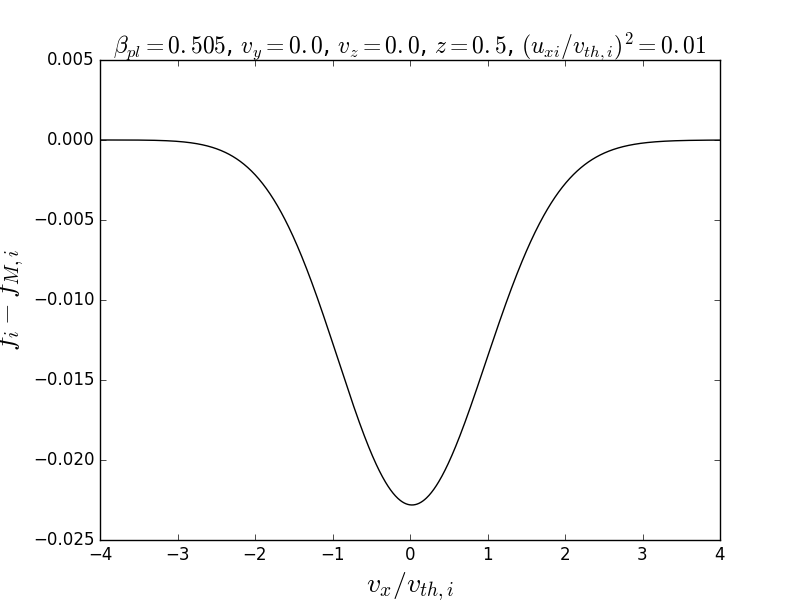}}}
\subfigure[]{\scalebox{0.32}{\includegraphics{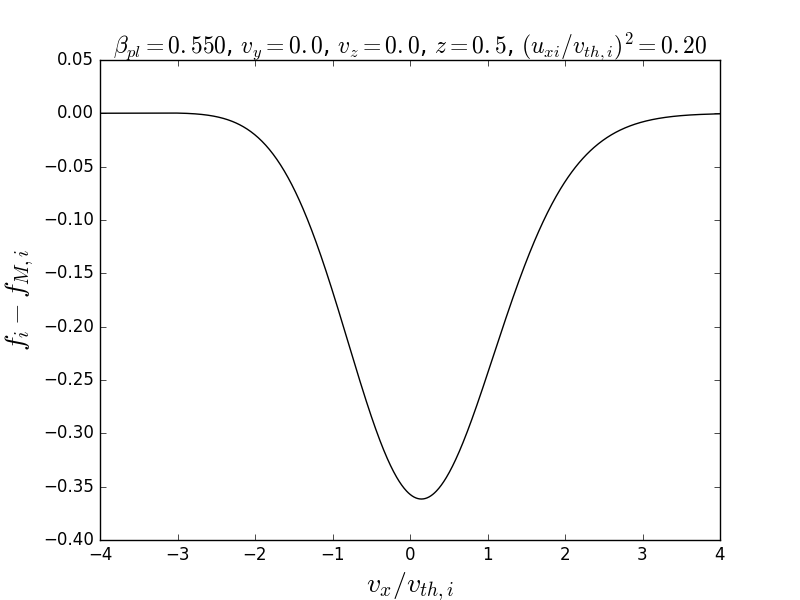}}}
\subfigure[]{\scalebox{0.32}{\includegraphics{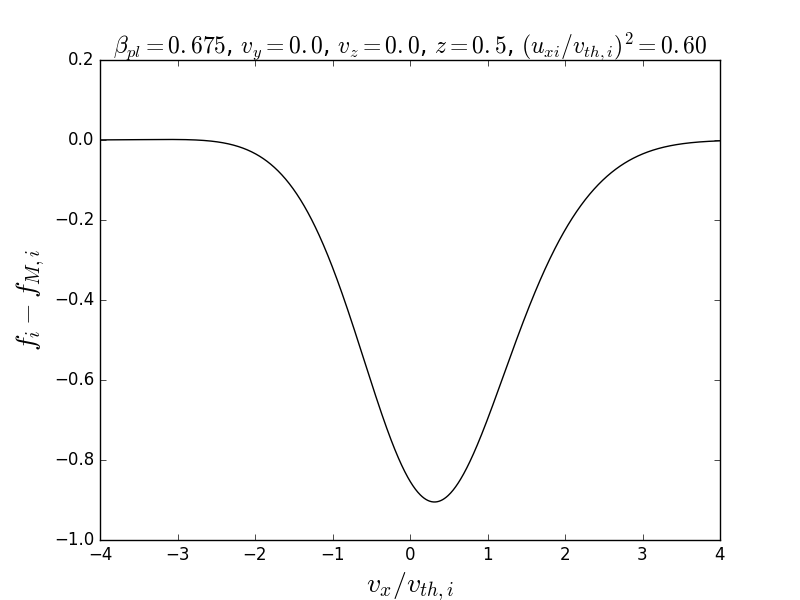}}}
\subfigure[]{\scalebox{0.32}{\includegraphics{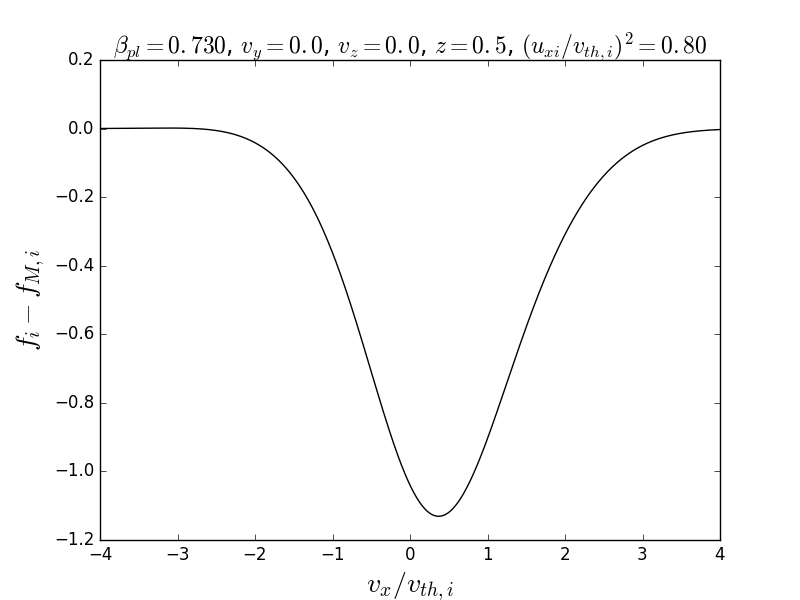}}}
\subfigure[]{\scalebox{0.32}{\includegraphics{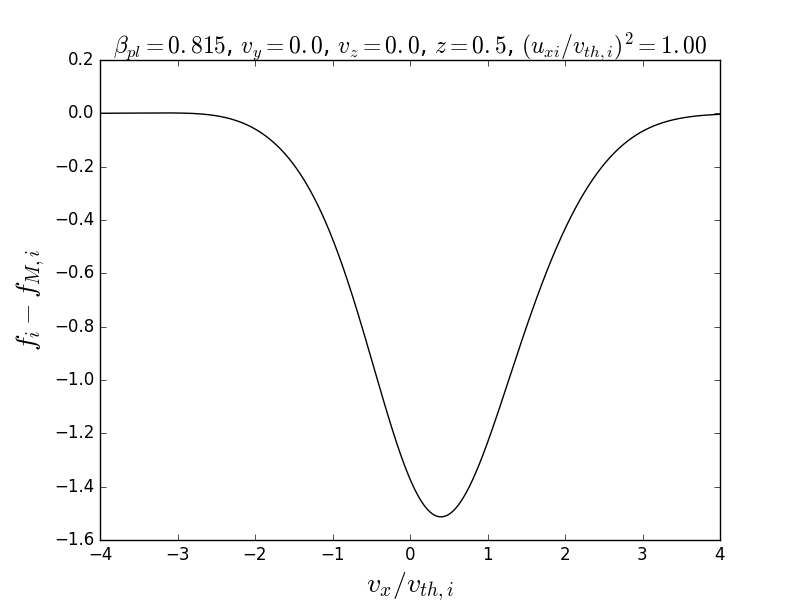}}}
\subfigure[]{\scalebox{0.32}{\includegraphics{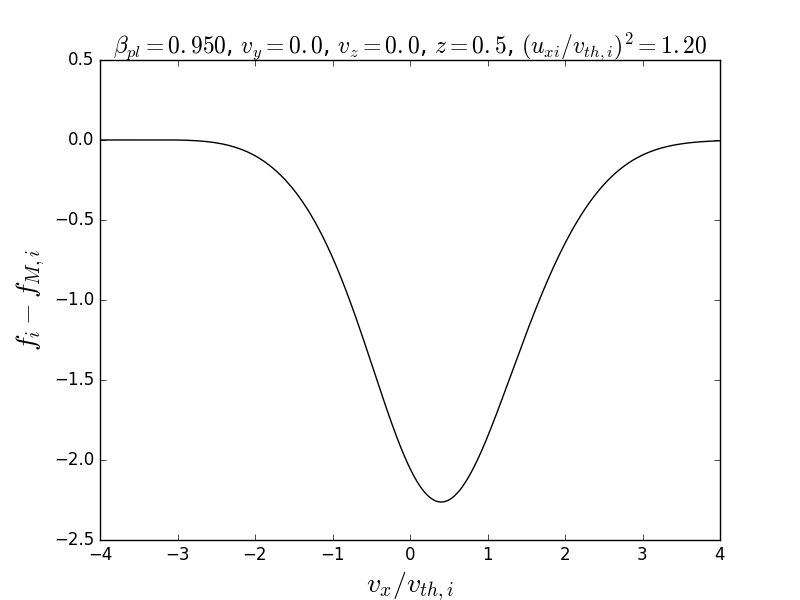}}}
\caption{Plots of $f_i-f_{M,i}$ in the $v_x$-direction (with $v_y=v_z$ and $z=0.5$) for various values of $u_{xi}/v_{th,i}$ and, hence, the plasma beta.}
\label{fig:quad_df_diff2}
\end{figure}

\begin{figure}
\centering\
\subfigure[]{\scalebox{0.32}{\includegraphics{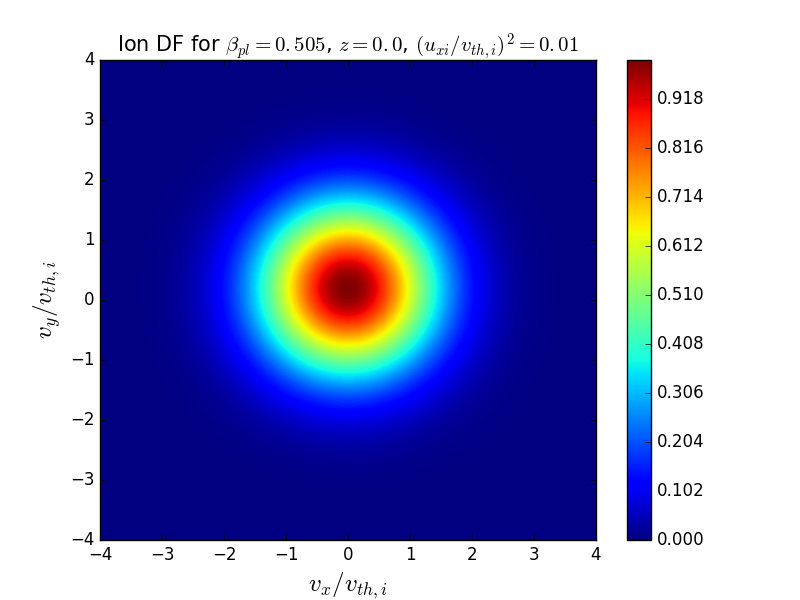}}}
\subfigure[]{\scalebox{0.32}{\includegraphics{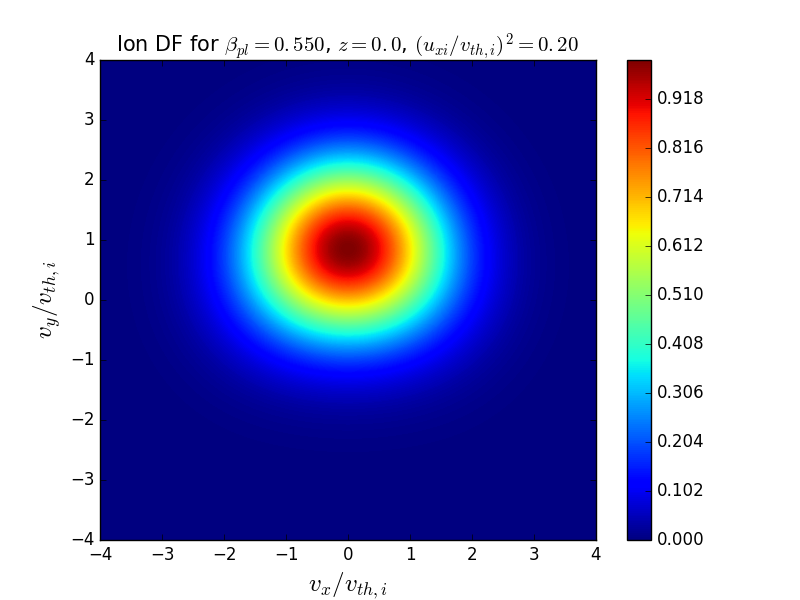}}}
\subfigure[]{\scalebox{0.32}{\includegraphics{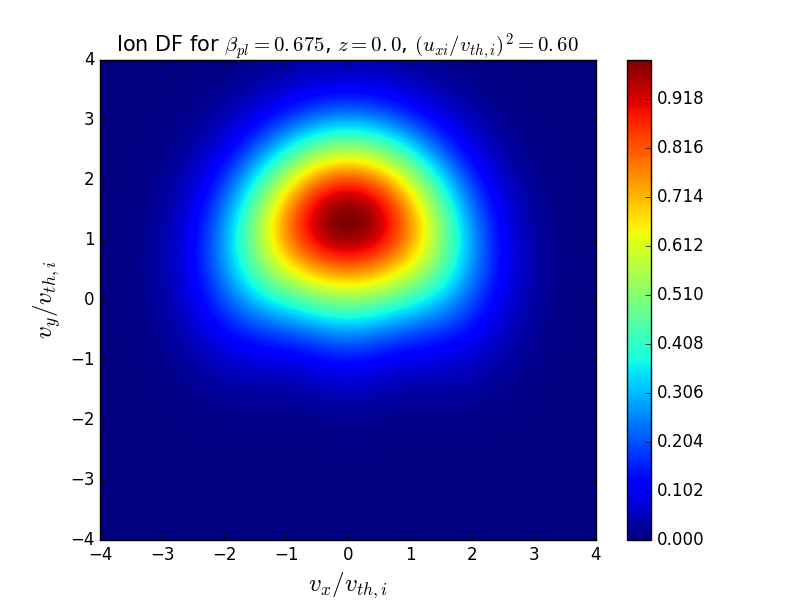}}}
\subfigure[]{\scalebox{0.32}{\includegraphics{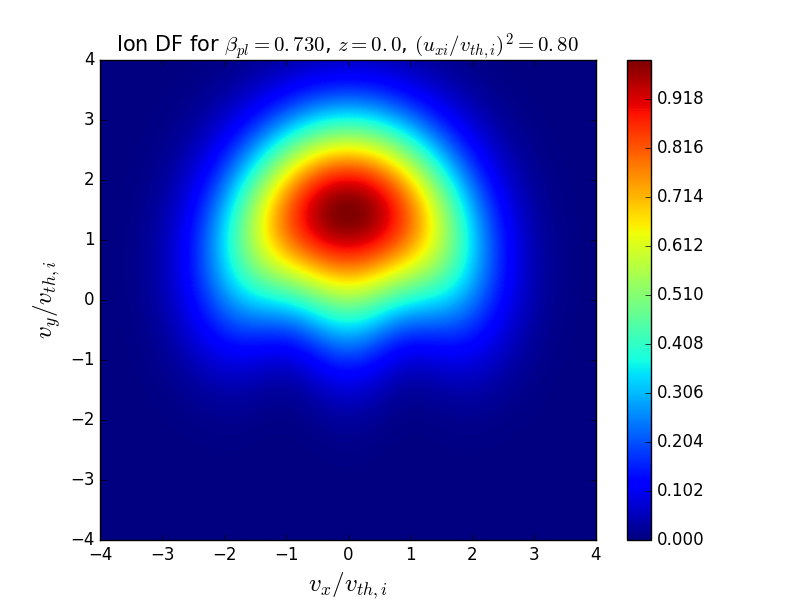}}}
\subfigure[]{\scalebox{0.32}{\includegraphics{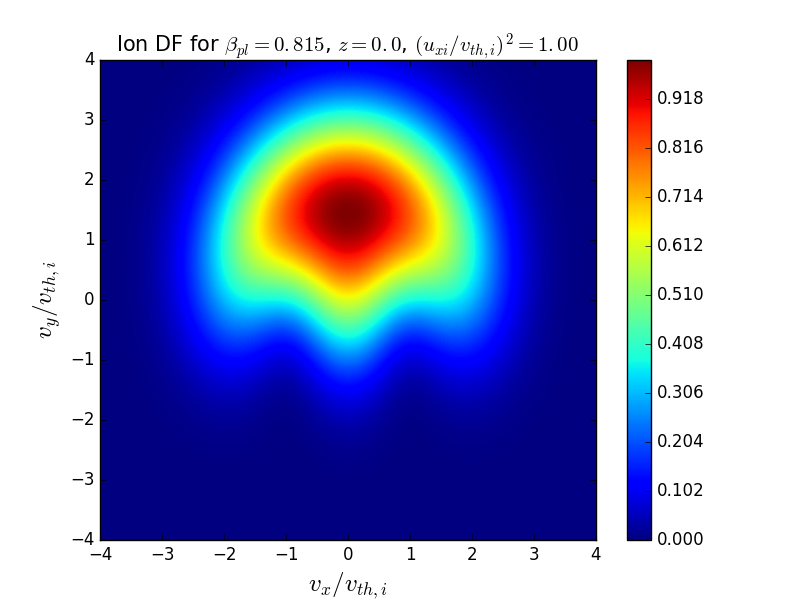}}}
\subfigure[]{\scalebox{0.32}{\includegraphics{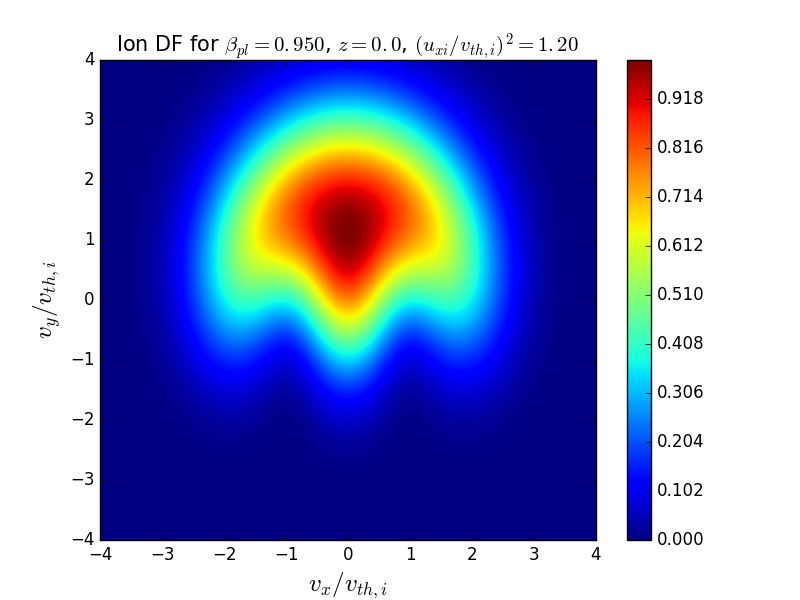}}}
\caption{Contour plots of the ion DF (\ref{quad_df2}) in the $v_x$-$v_y$-plane ($z=0$) for various values of $u_{xi}/v_{th,i}$ and, hence, the plasma beta. The DFs are normalised to have a maximum value of one in each case.}
\label{fig:quad_df_contour}
\end{figure}

\begin{figure}
\centering\
\subfigure[]{\scalebox{0.32}{\includegraphics{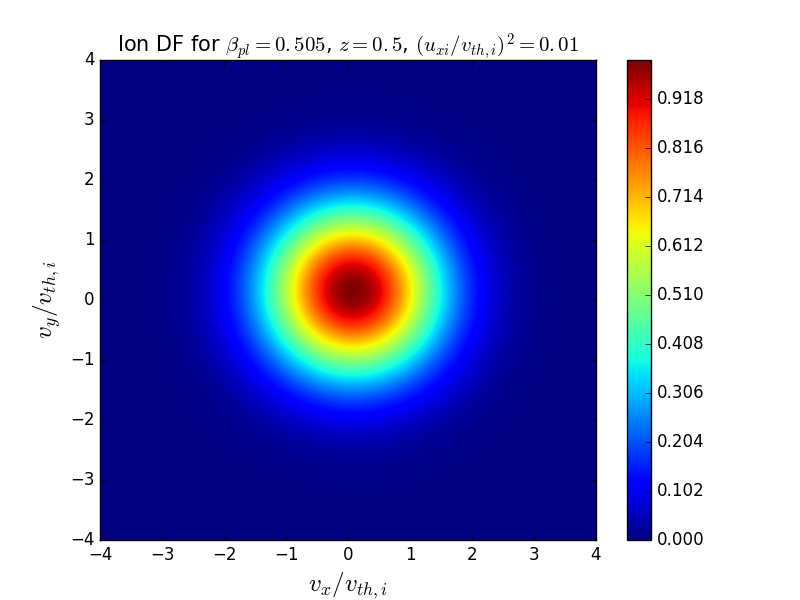}}}
\subfigure[]{\scalebox{0.32}{\includegraphics{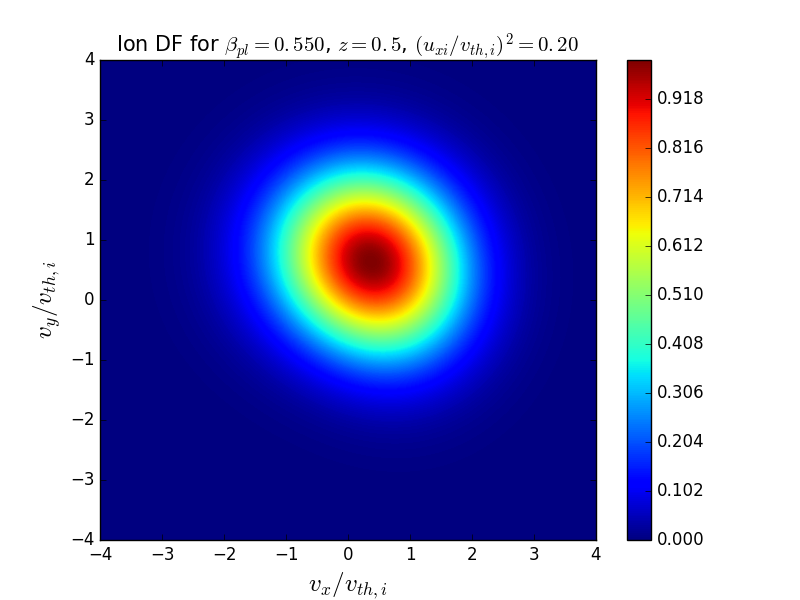}}}
\subfigure[]{\scalebox{0.32}{\includegraphics{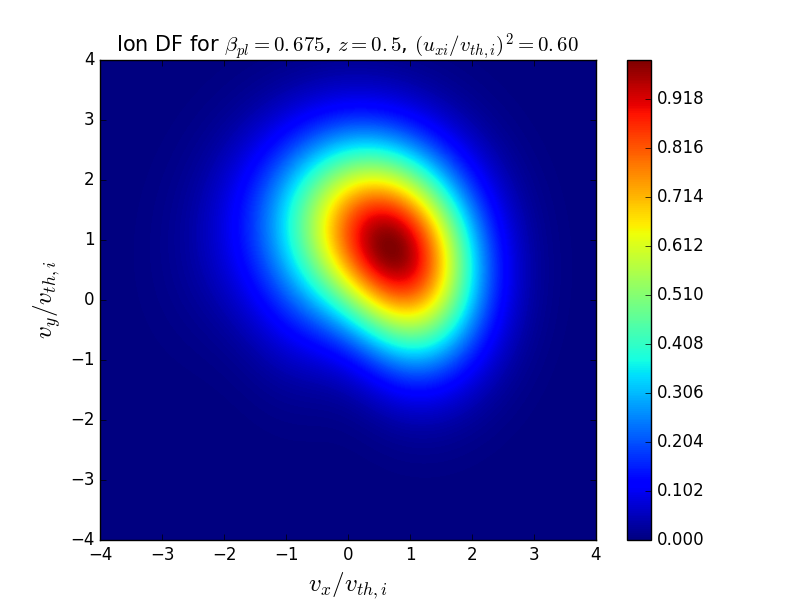}}}
\subfigure[]{\scalebox{0.32}{\includegraphics{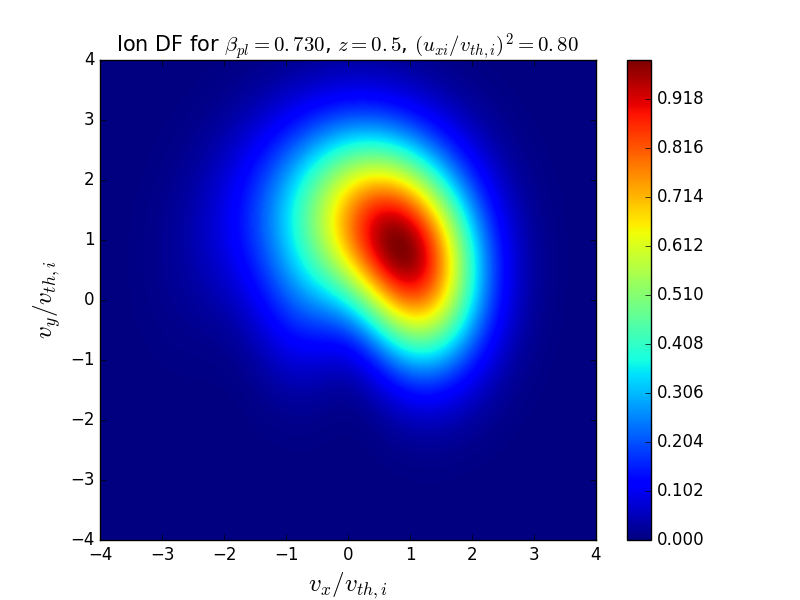}}}
\subfigure[]{\scalebox{0.32}{\includegraphics{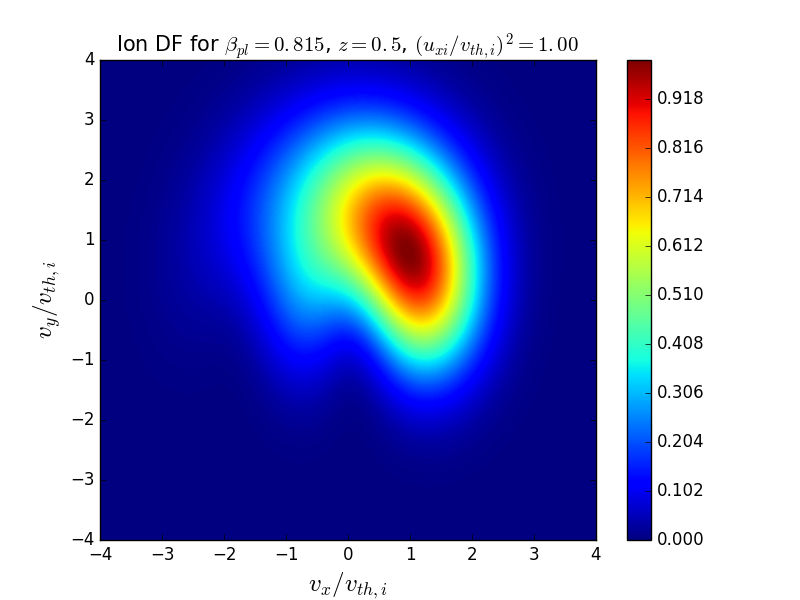}}}
\subfigure[]{\scalebox{0.32}{\includegraphics{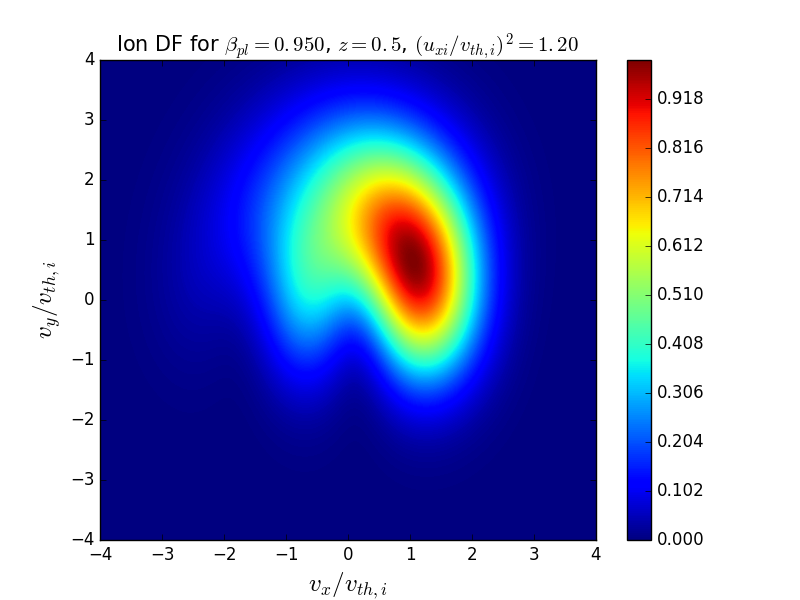}}}
\caption{Contour plots of the ion DF (\ref{quad_df2}) in the $v_x$-$v_y$-plane ($z=0.5$) for various values of $u_{xi}/v_{th,i}$ and, hence, the plasma beta. The DFs are normalised to have a maximum value of one in each case.}
\label{fig:quad_df_contour2}
\end{figure}

\begin{figure}
\centering\
\subfigure[]{\scalebox{0.32}{\includegraphics{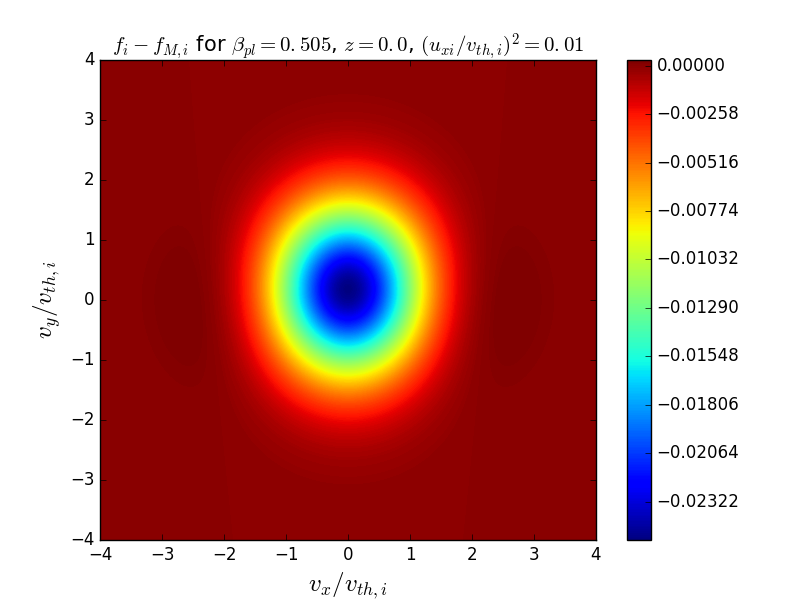}}}
\subfigure[]{\scalebox{0.32}{\includegraphics{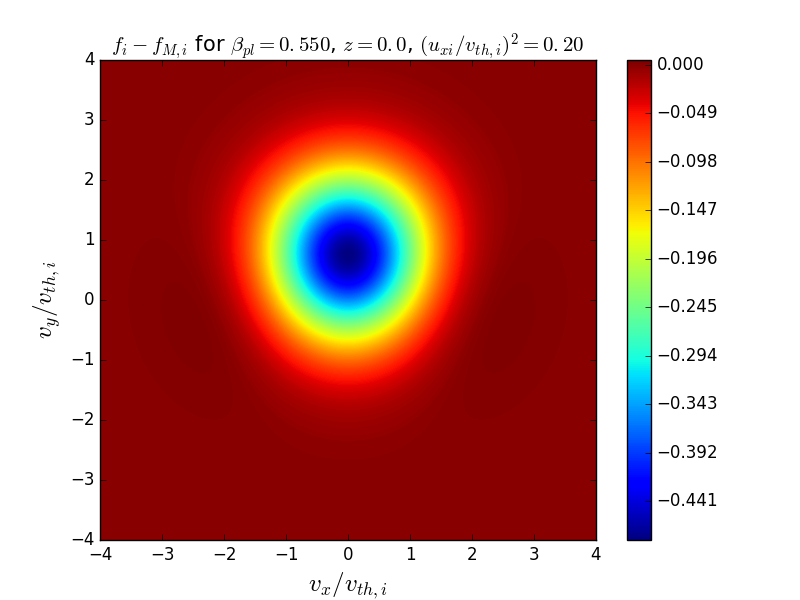}}}
\subfigure[]{\scalebox{0.32}{\includegraphics{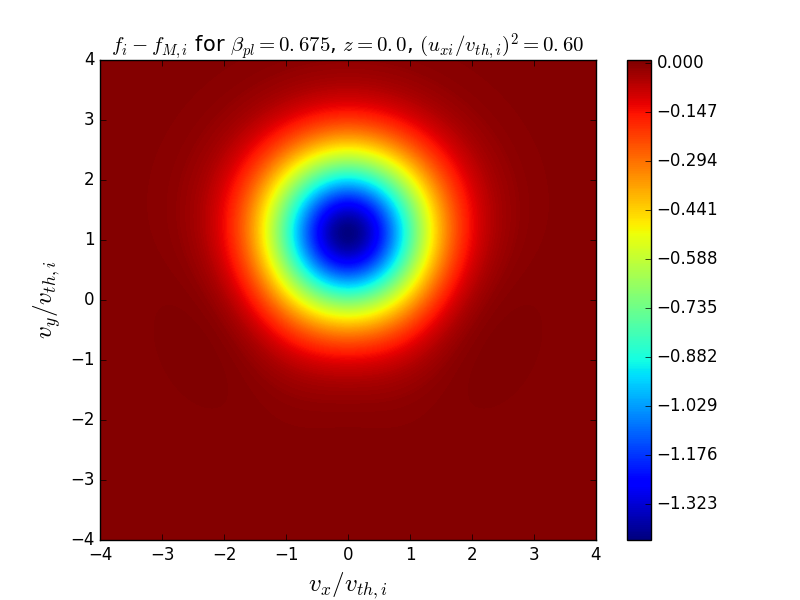}}}
\subfigure[]{\scalebox{0.32}{\includegraphics{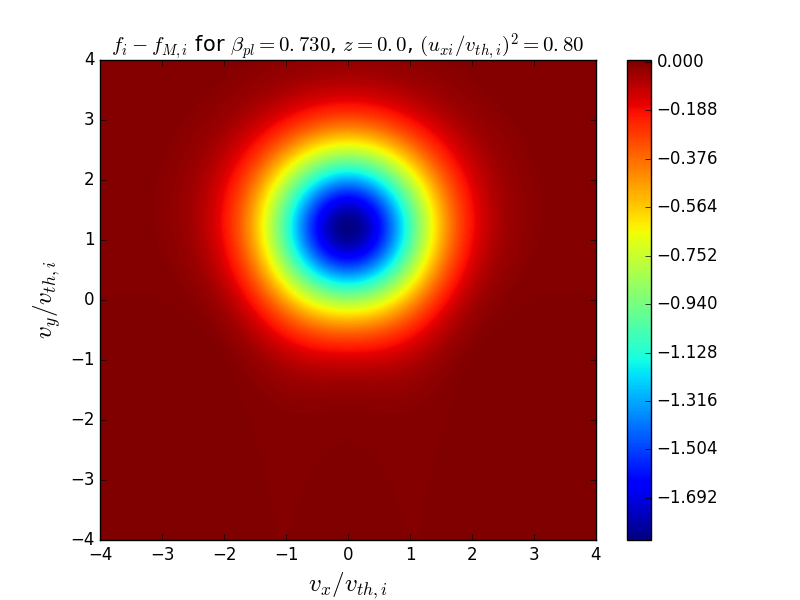}}}
\subfigure[]{\scalebox{0.32}{\includegraphics{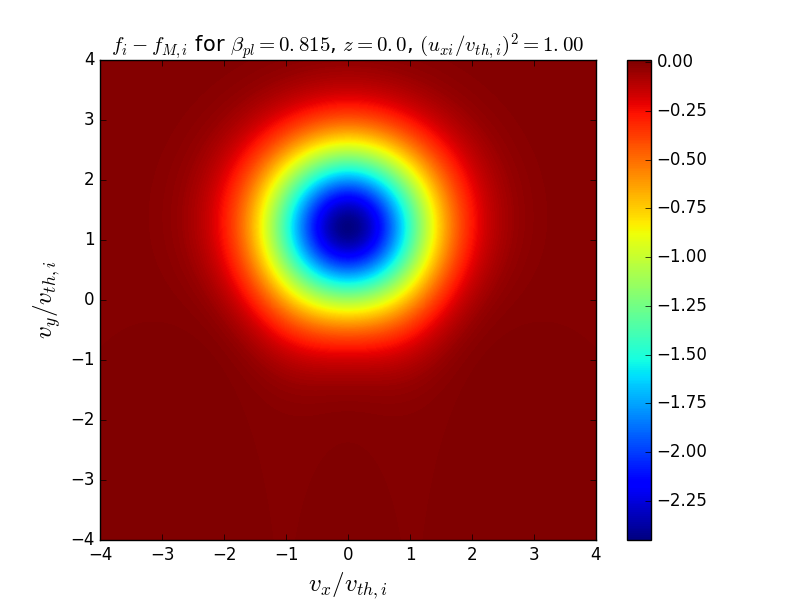}}}
\subfigure[]{\scalebox{0.32}{\includegraphics{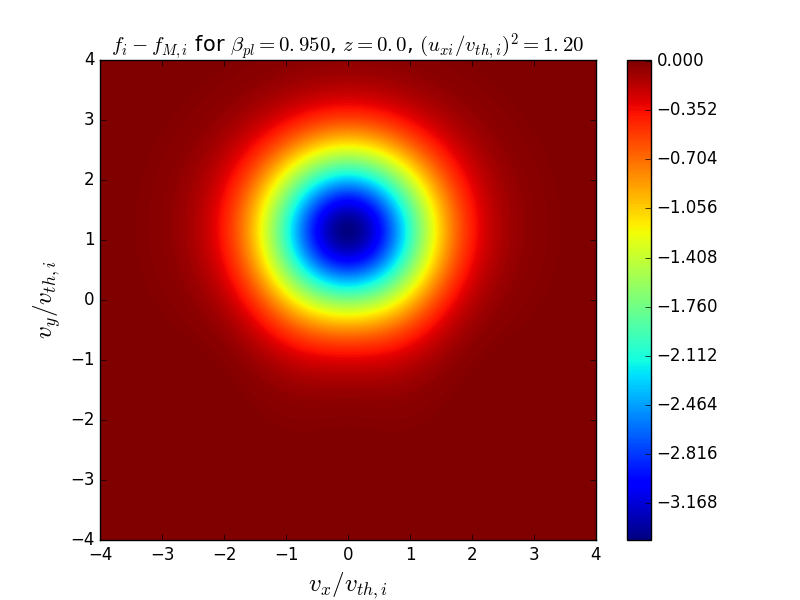}}}
\caption{Contour plots of $f_i-f_{M,i}$ in the $v_x$-$v_y$-plane ($z=0$) for the parameter values from Figure \ref{fig:quad_df_contour}.}
\label{fig:quad_df_contour_diff}
\end{figure}

\begin{figure}
\centering\
\subfigure[]{\scalebox{0.32}{\includegraphics{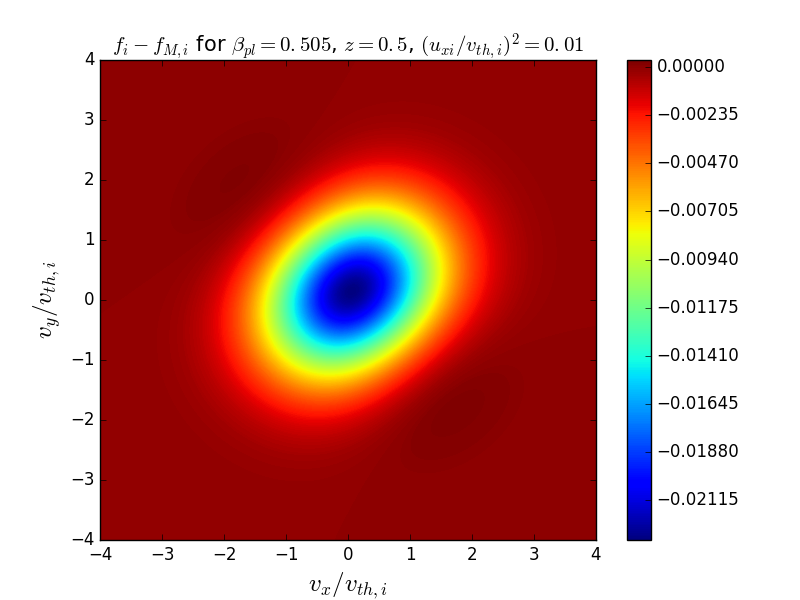}}}
\subfigure[]{\scalebox{0.32}{\includegraphics{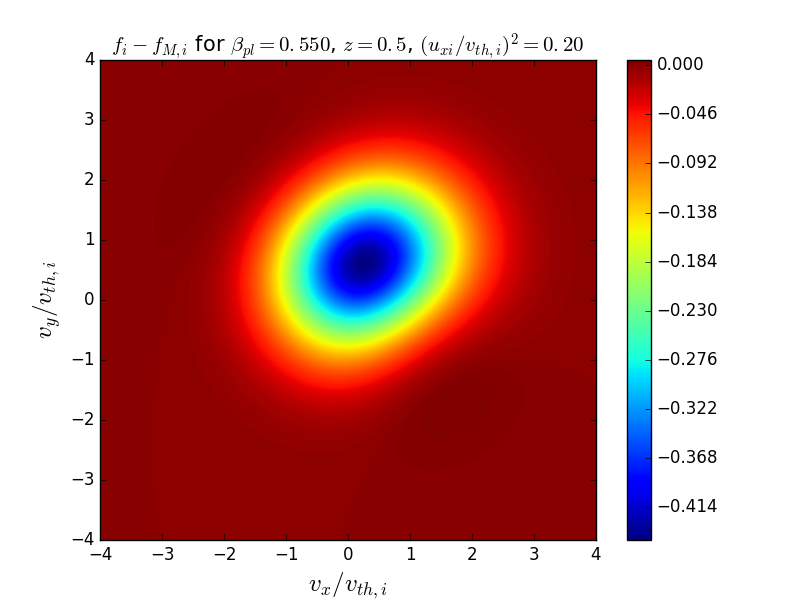}}}
\subfigure[]{\scalebox{0.32}{\includegraphics{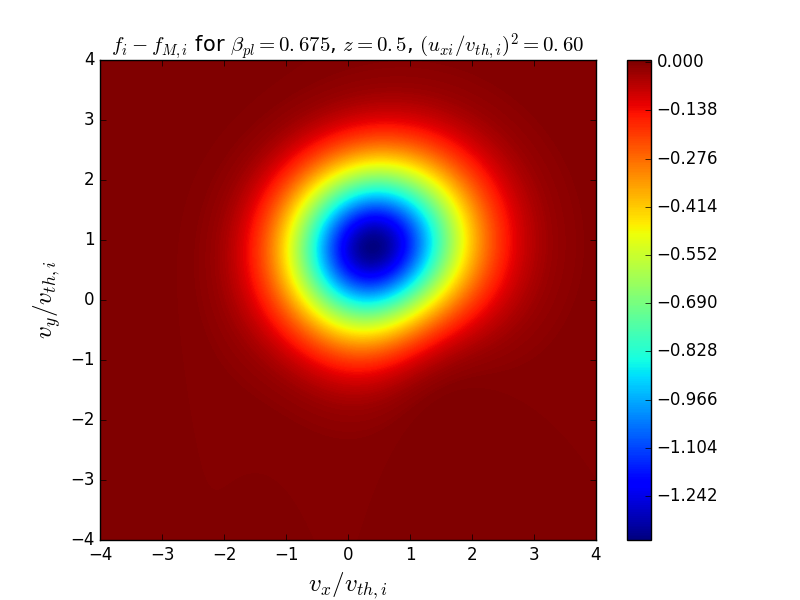}}}
\subfigure[]{\scalebox{0.32}{\includegraphics{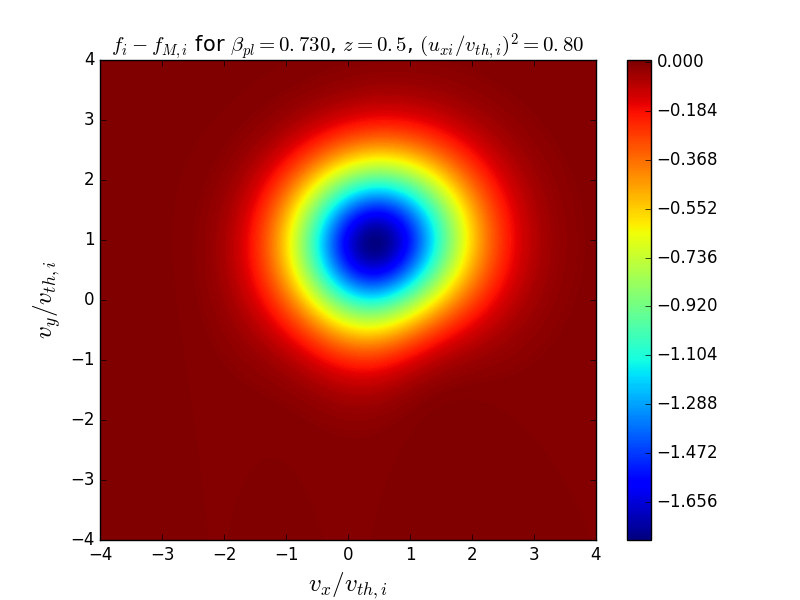}}}
\subfigure[]{\scalebox{0.32}{\includegraphics{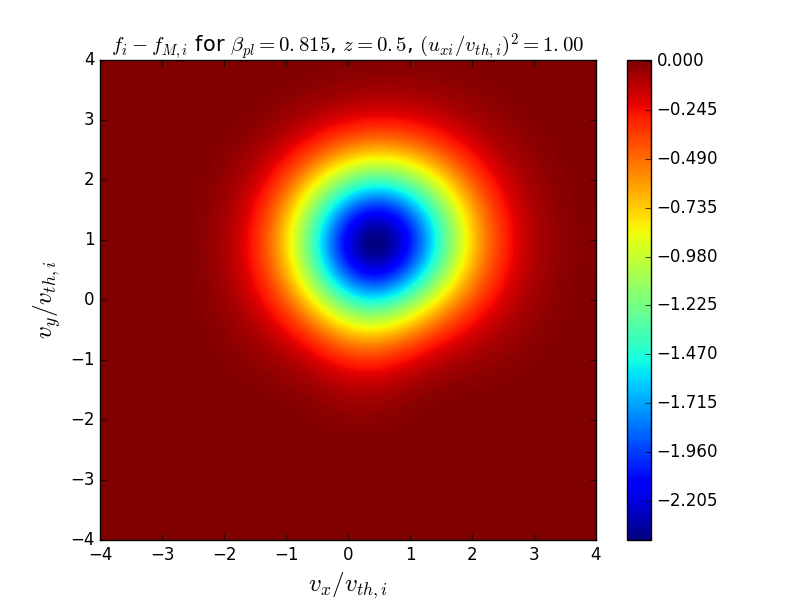}}}
\subfigure[]{\scalebox{0.32}{\includegraphics{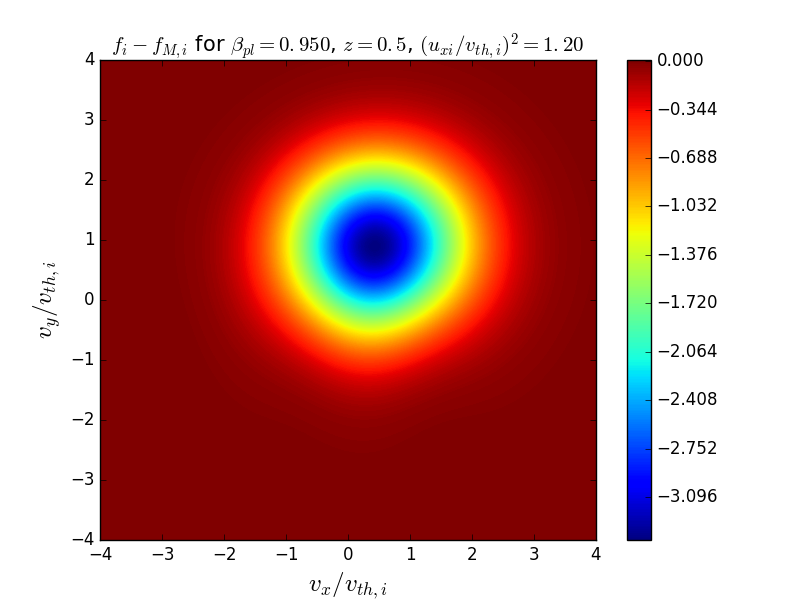}}}
\caption{Contour plots of $f_i-f_{M,i}$ in the $v_x$-$v_y$-plane ($z=0.5$) for the parameter values from Figure \ref{fig:quad_df_contour}.}
\label{fig:quad_df_contour_diff2}
\end{figure}

\begin{figure}
\centering\
\subfigure[]{\scalebox{0.32}{\includegraphics{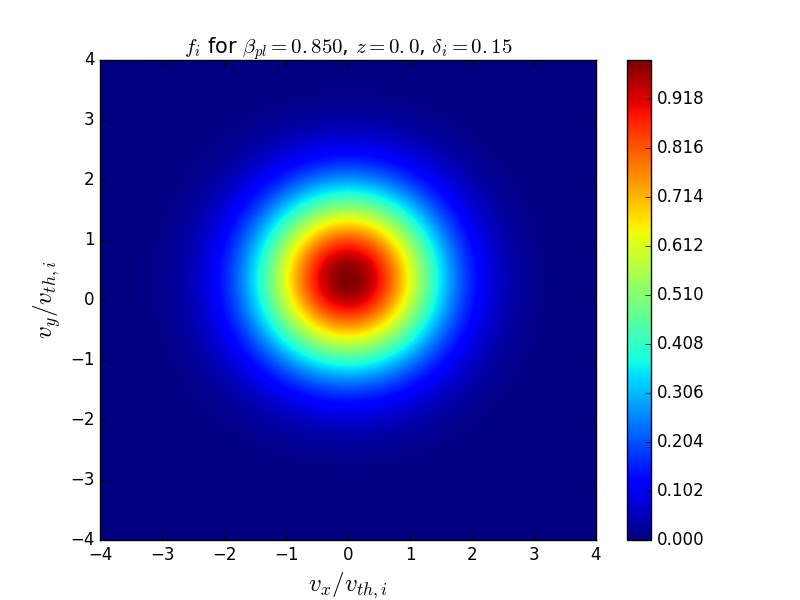}}}
\subfigure[]{\scalebox{0.32}{\includegraphics{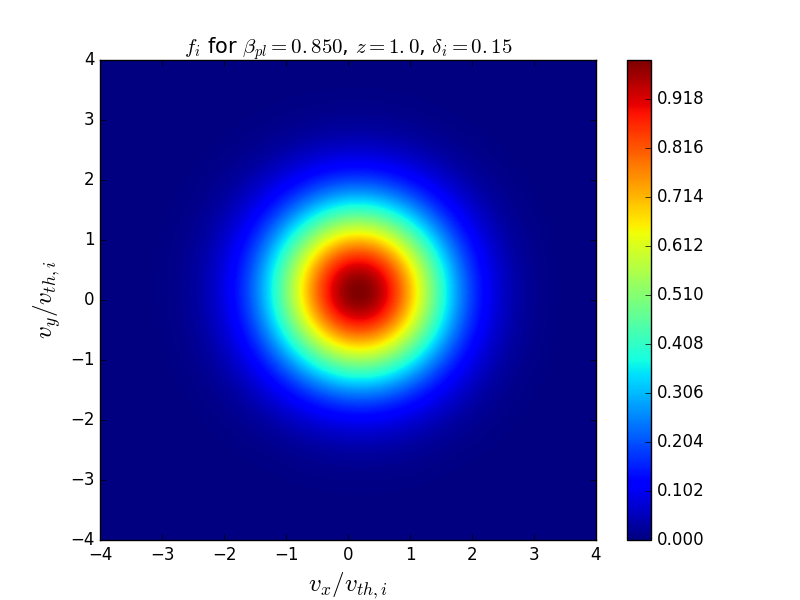}}}
\caption{Contour plot of the ion DF (\ref{quad_df2})  in the $v_x$-$v_y$-plane ($v_z=0$) for the parameters used by \cite{Allanson-2015}, for (a) $z=0$ and (b) $z=1.0$.}
\label{fig:compare1}
\end{figure}

\begin{figure}
\centering\
\subfigure[]{\scalebox{0.32}{\includegraphics{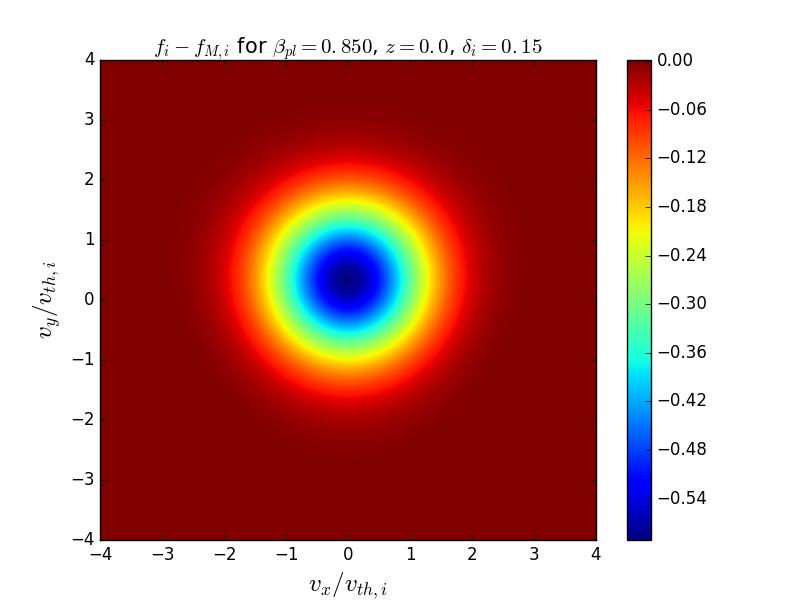}}}
\subfigure[]{\scalebox{0.32}{\includegraphics{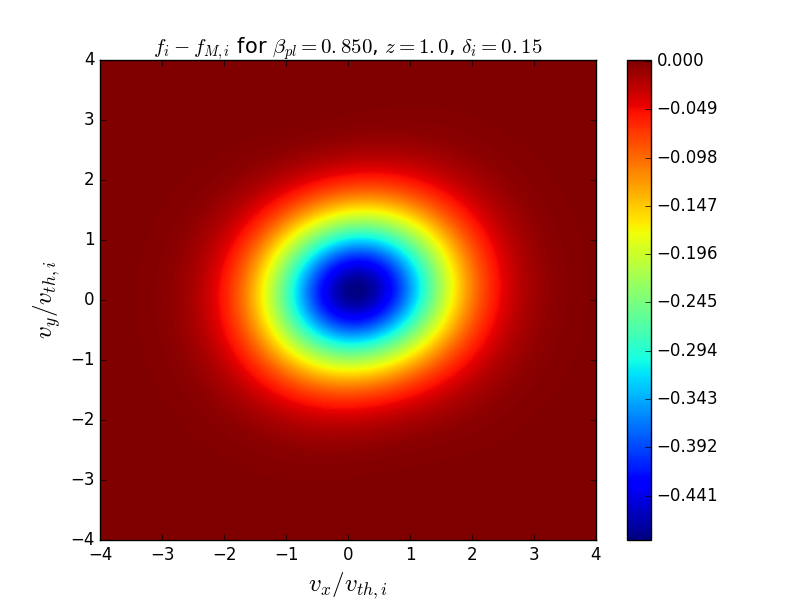}}}
\caption{Contour plot of $f_{i}-f_{M,i}$ in the $v_x$-$v_y$-plane ($v_z=0$) for the parameters used by \cite{Allanson-2015}, for (a) $z=0$ and (b) $z=1.0$.}
\label{fig:compare2}
\end{figure}

\begin{figure}
\centering\
\subfigure[]{\scalebox{0.32}{\includegraphics{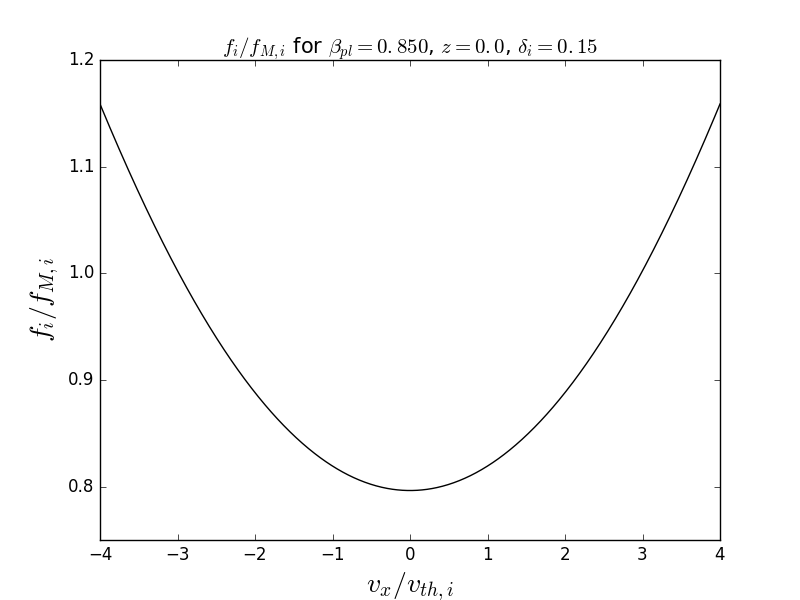}}}
\subfigure[]{\scalebox{0.32}{\includegraphics{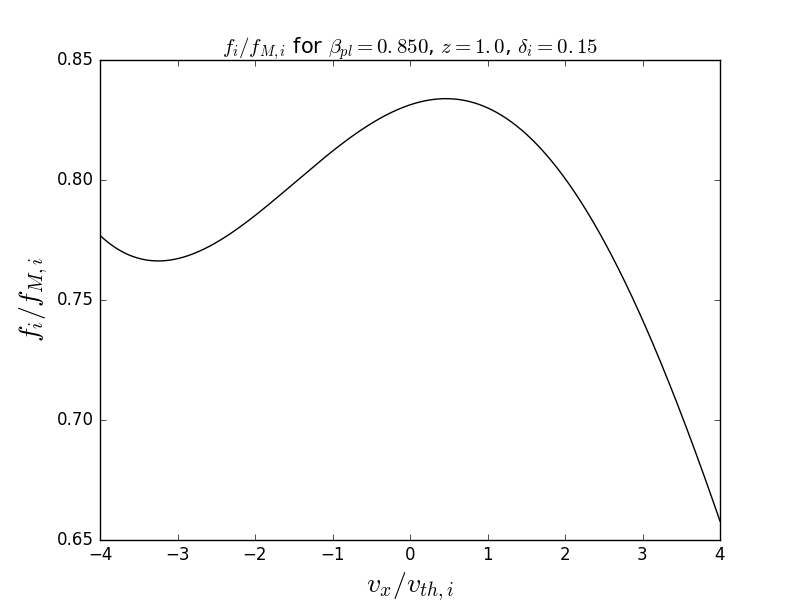}}}
\caption{Contour plot of $f_{i}/f_{M,i}$ in the $v_x$-direction for the parameters used by \cite{Allanson-2015}, for (a) $z=0$ and (b) $z=1.0$.}
\label{fig:compare3}
\end{figure}

\begin{figure}
\centering\
\subfigure[]{\scalebox{0.32}{\includegraphics{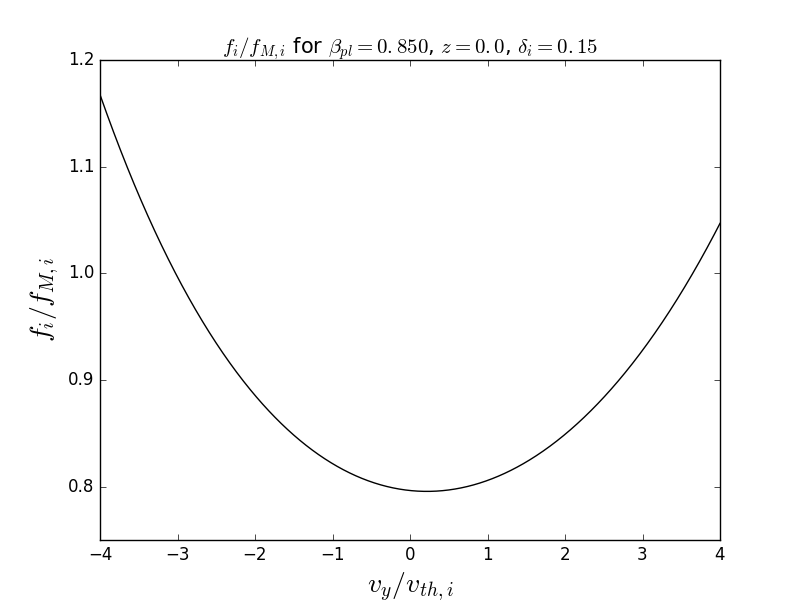}}}
\subfigure[]{\scalebox{0.32}{\includegraphics{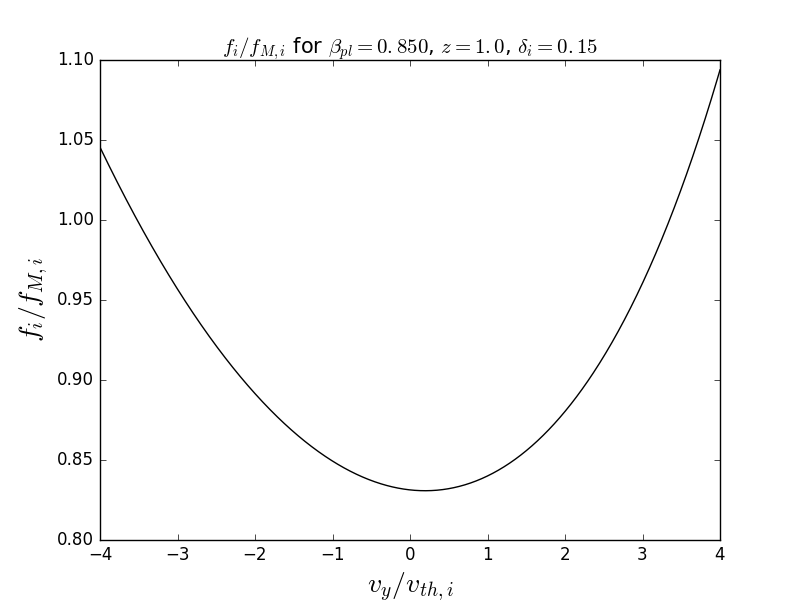}}}
\caption{Contour plot of $f_{i}/f_{M,i}$ in the $v_y$-direction for the parameters used by \cite{Allanson-2015}, for (a) $z=0$ and (b) $z=1.0$.}
\label{fig:compare4}
\end{figure}

\begin{figure}
\centering\
\subfigure[]{\scalebox{0.32}{\includegraphics{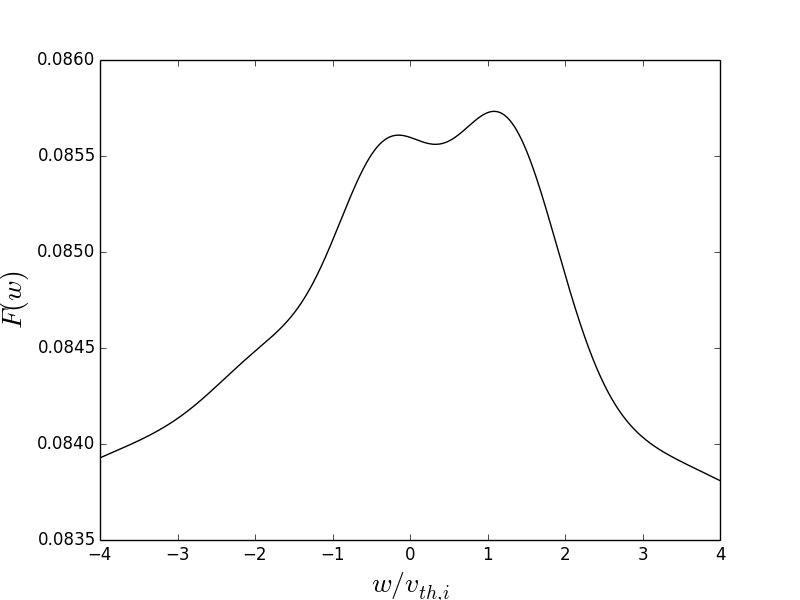}}}
\subfigure[]{\scalebox{0.32}{\includegraphics{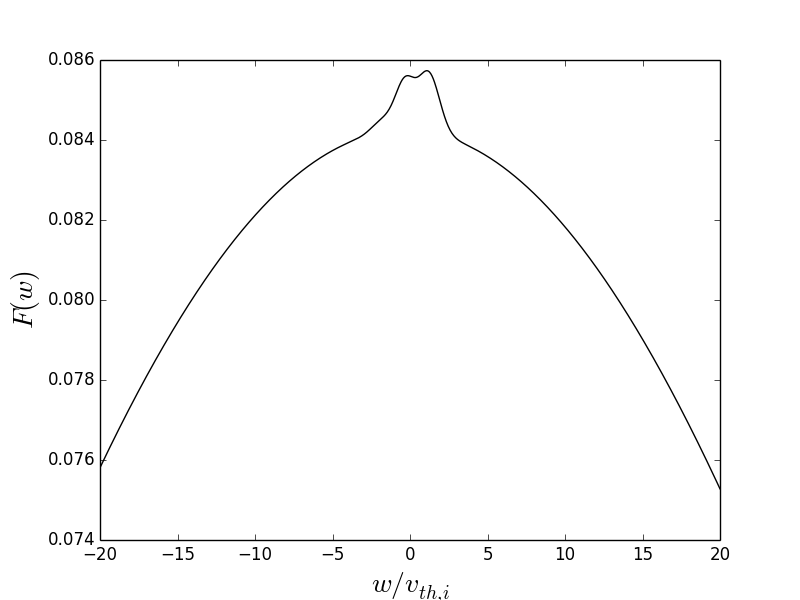}}}
\subfigure[]{\scalebox{0.32}{\includegraphics{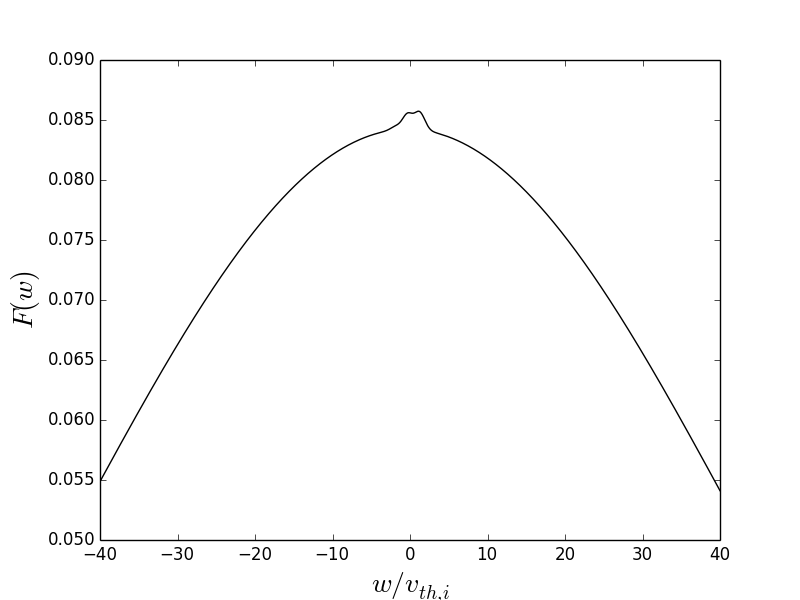}}}
\subfigure[]{\scalebox{0.32}{\includegraphics{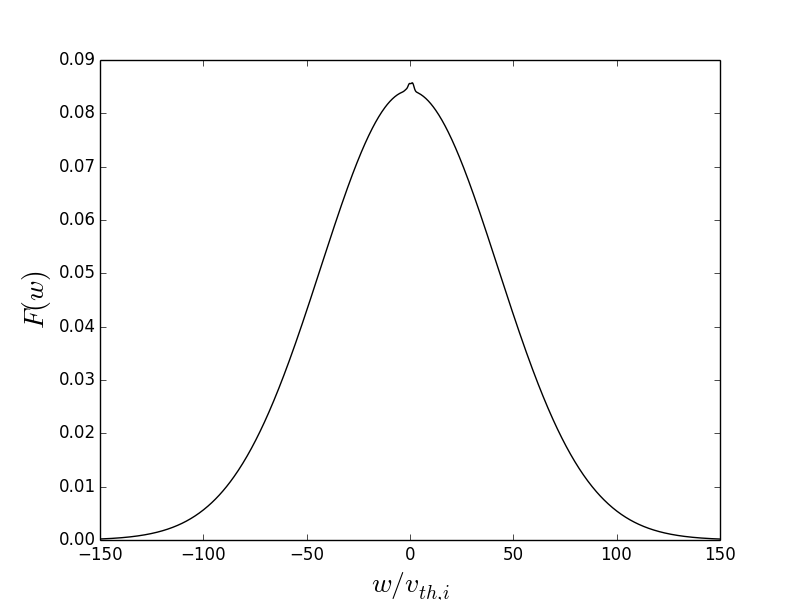}}}
\caption{Plots of $F(w)$ (normalised to $n_0/(\sqrt{2\pi}v_{th,i})$) over four different velocity ranges, for the parameters $\beta_{pl}=0.95$, $(u_{xi}/v_{th,i})^2=1.2$, $\theta=\pi/2$, $\phi=0$, $z=0.3$ and $T_i/T_e=1$.}
\label{fig:penrose}
\end{figure}

\end{document}